\documentclass[11pt,a4paper]{article}
\usepackage{jcappub}
\usepackage{color}
\usepackage{amsfonts, amssymb,amsmath}
\usepackage{slashed}
\usepackage{braket}
\usepackage[normalem]{ulem}
\usepackage{comment}
\usepackage{xspace} 
\usepackage{siunitx}
\usepackage{multirow}
\usepackage{subcaption}
\usepackage{placeins}
\usepackage{afterpage}
\usepackage{booktabs}

\newcommand{\fnlloc}{f_{\rm NL}^{\rm loc}}
\newcommand{\bx}{\mathbf{x}}

\newcommand{\bs}{\mathbf{s}}

\newcommand{\fnleq}{f_{\rm NL}^{\rm equi}}

\newcommand{\eos}{\textsc{Eos}\xspace}
\newcommand{\dataset}{\textsc{Dataset}\xspace}

\title{Fisher Forecasts for Primordial non-Gaussianity from Persistent Homology}

\author[a,b,c,d]{Matteo Biagetti,}
\author[e]{Juan Calles,}
\author[e]{Lina Castiblanco,}
\author[f]{Alex Cole}
\author[e]{and Jorge Nore\~na.}

\affiliation[a]{Institute for Fundamental Physics of the Universe, Via Beirut 2, 34151 Trieste, Italy}
\affiliation[b]{SISSA - International School for Advanced Studies, Via Bonomea 265, 34136 Trieste,  Italy}
\affiliation[c]{Istituto Nazionale di Astrofisica, Osservatorio Astronomico di Trieste, via Tiepolo 11, 34143 Trieste, Italy}
\affiliation[d]{Istituto Nazionale di Fisica Nucleare, Sezione di Trieste,  via  Valerio  2,  34127 Trieste,  Italy}
\affiliation[e]{Instituto de F\'isica, Pontificia Universidad Cat\'olica de Valpara\'iso, Casilla 4950, Valpara\'iso, Chile}
\affiliation[f]{GRAPPA and ITFA, Institute of Physics, University of Amsterdam, Science Park 904, 1090 GL Amsterdam, the Netherlands}

\abstract{	We study the information content of summary statistics built from the multi-scale topology of large-scale structures on primordial non-Gaussianity of the local and equilateral type.
	We use halo catalogs generated from numerical N-body simulations of the Universe on large scales as a proxy for observed galaxies. Besides calculating the Fisher matrix for halos in real space, we also check more realistic scenarios  in redshift space.
	Without needing to take a distant observer approximation, we 
	place the observer on a corner of the box.
	We also add redshift errors mimicking spectroscopic and photometric samples. We perform several tests to assess the reliability of our Fisher matrix, including the Gaussianity of our summary statistics and convergence. We find that the marginalized 1-$\sigma$ uncertainties in redshift space are $\Delta \fnlloc \sim 16$ and $\Delta \fnleq \sim 41 $ on a survey volume of $1$ $($Gpc$/h)^3$. These constraints are weakly affected by redshift errors. We close by speculating as to how this approach can be made  robust against small-scale uncertainties by exploiting (non)locality.}

\begin{document}

\maketitle

\flushbottom 

\newpage

\section{Introduction}

Primordial perturbations are, to a very good approximation, Gaussian distributed, as recently constrained by Cosmic Microwave Background (CMB) experiments \cite{Akrami:2018odb} and galaxy surveys \cite{Castorina:2019wmr,Cabass:2022wjy, DAmico:2022gki}. An even small deviation from Gaussianity has important implications  for our understanding of the physics during inflation, revealing interactions among fields at very high energy (see \cite{Bartolo:2004if,Chen:2010xka} for reviews). Several experimental efforts will constrain these interactions with unprecedented accuracy in the near future (see \cite{Meerburg:2019qqi} for a recent overview). While the constraining power of the CMB on primordial non-Gaussianity is relatively simple to predict, the same cannot be said about large scale structures (LSS) observations. The challenge is  due to the fact that the cosmic web is contaminated by non-Gaussianities that are not primordial, but instead produced by gravitational instability. Yet, the signal-to-noise of galaxy surveys and intensity mapping experiments typically grows as $k_{\rm max}^3 \times V_{\rm survey}$ as compared to $\ell^2_{\rm max}$ for CMB. Hence, LSS surveys have a strong potential to dominate constraints on primordial non-Gaussianity and inflationary models in general in the near future.\footnote{A first example in this direction was achieved already with BOSS DR12 galaxies in \cite{Beutler:2019ojk}. The analysis of the galaxy power spectrum provided improved constraints on inflationary models with primordial features over constraints from Planck.}

More broadly, a central problem in modern cosmology is the extraction of maximum information content regarding the universe's initial conditions and dynamics from observations. Beyond the traditional program of low-order correlation functions in Fourier space, there is an ongoing effort to characterize cosmological information content of real-space and higher-order statistics such as the $k$-nearest neighbor distribution \cite{Banerjee:2020umh,Banerjee:2021cmi,Banerjee:2021hkg} and one-point PDF \cite{Mao:2014caa,Uhlemann:2017tex,Nusser:2018vym,Friedrich:2019byw,Uhlemann:2019gni}. Several of the present authors recently proposed a new class of statistics derived from computational topology for this purpose \cite{Biagetti:2020skr}. These statistics describe the multi-scale arrangement of data into clusters, loops, and voids, formalized by the theory of persistent homology (see \cite{zomorodian2009computational,edelsbrunner2010computational} for general references). The theory of persistent homology has been fruitfully applied to sensor network analysis \cite{desilva2007}, image analysis \cite{carlsson2008local}, virology \cite{chan2013topology}, protein structure \cite{Gameiro2015}, string theory \cite{Cole:2018emh}, statistical physics \cite{Cole:2020hjx}, and more. Over the last several years, persistent homology has begun to see use in quantitatively constraining cosmology \cite{Cole:2017kve,Heydenreich:2020hrr,Biagetti:2020skr}. In the restricted context of dark matter-only simulations in a box, \cite{Biagetti:2020skr} found sensitivity to local primordial non-Gaussianity competitive with the scale-dependent bias, while relying on complementary information (i.e.\ the higher-order position-space arrangement of dark matter halos on mildly nonlinear scales rather than the low-order information in momentum space at large wavelength). In some sense, local primordial non-Gaussianity is special: even without resorting to high-order correlation function information, future surveys are already expected to significantly improve CMB constraints \cite{Dore:2014cca}. This is thanks to the fact that local primordial non-Gaussianity produces a modulation of short-scale modes by long-wavelength modes of the gravitational potential and its spatial derivatives, and therefore an effect of enhancement/suppression of the tracer power spectrum at large scales, where late non-Gaussianities are negligible \cite{Dalal:2007cu,Slosar:2008hx,Matarrese:2008nc} (see \cite{Biagetti:2019bnp} for a recent review).

For other types of primordial non-Gaussianity there is no such an advantage.
Detection of these non-Gaussianities would have interesting theoretical implications. For instance, a detection of equilateral non-Gaussianity at the level of $\fnleq\gtrsim 1$ would exclude slow-roll inflation as a viable inflationary scenario \cite{Baumann:2014cja,Baumann:2015nta}. 
Furthermore, if gravitational waves generated during inflation are ever observed, constraints of the order of $\Delta \fnleq \sim 1$ can be used to tell whether they were sourced by the vacuum or other fields (see e.g. \cite{Mirbabayi:2014jqa, Ferreira:2014zia}). 

However, the outlook for $\fnleq$ has been mostly grim. Recent measurement of the three-point function of galaxies in BOSS using state of the art modeling has shown that limits are of the order of a few hundreds for $\fnleq$ \cite{Cabass:2022wjy,DAmico:2022gki}.  
Pessimism for $\fnleq$ is largely motivated by analyses based on low-order correlation functions in momentum space. 
When one restricts to these summary statistics, the primordial signal is degenerate with unknown aspects of nonlinear evolution, e.g.\ galaxy biasing. However, as recently explained in \cite{Baumann:2021ykm}, one can do significantly better by considering survey data in position space, i.e.\ at the map level. The physical argument in \cite{Baumann:2021ykm} goes as follows. Uncertainties in gravitational evolution are constrained by locality, and therefore these can affect map-level information only on scales below some $R_*$. On the other hand, primordial non-Gaussianity gives rise to non-local effects at late times, as local interactions during inflation have been stretched to super-horizon scales. Therefore there exist observables that trace primordial non-Gaussianity but remain insensitive to physics below $R_*$. These observables are manifestly protected against small-scale uncertainties. 

It is therefore of interest to characterize the information content of real-space statistics regarding primordial non-Gaussianity. In this paper, we perform this characterization for topological observables. Since topological features have extent, they can trace nonlocal correlations such as those generated by primordial non-Gaussianity. On the other hand, as we elaborate in Sec.\ \ref{sec:PHreview}, the features we study are grown locally by coarse-graining the halo distribution. In this sense, there is a cutoff we may apply in order to ignore irrelevant, i.e. small-scale, features. In this work, we find that simple, low-dimensional binned distributions of the linking scale at which a feature forms and the linking scale at which a feature is erased provide significant information regarding primordial non-Gaussianity. We estimate this information via a Fisher matrix formalism for N-body simulations, forecasting constraints of $\Delta \fnlloc \sim 15$, $\Delta \fnleq\sim 40$ for a survey volume of 1 (Gpc/$h)^3$. We begin by considering halos in position space. We then move to redshift space, considering observers both in the plane parallel approximation and in the wide angle case, finding that the constraints are not significantly degraded. We find similar robustness when redshift errors are included. We interpret this robustness as a consequence of well-known stability properties of persistent homology \cite{cohen2007stability}. Therefore, we anticipate that, in the presence of a parameterized model of small-scale physics such as galaxy formation, persistent homology may provide a ``manifestly protected'' observable along the lines of \cite{Baumann:2021ykm}.

The rest of the paper is organized as follows. In Sec.\ \ref{sec:PHreview} we review necessary aspects of persistent homology, including the coarse-graining scheme we implement and the summary statistics we employ. In Sec.\ \ref{sec:phhalo} we describe in more detail local and equilateral non-Gaussianity, and how they can be implemented in N-body simulations. In Sec.\ \ref{sec:implementation} we outline our methodology for computing the Fisher information, including our simulation set and implementation of redshift errors. In Sec.\ \ref{sec:results} we give the results of our analysis. In Sec.\ \ref{sec:small} we explore the robustness of our topological analysis to small scale effects. In Sec.\ \ref{sec:conclusions} we conclude.

\section{Formalism}\label{sec:PHreview}
Here we briefly outline the construction and interpretation of our summary statistics. For a more complete presentation on applying persistent homology to a cosmological setup, we refer to \cite{Biagetti:2020skr}. For general references on persistent homology, see \cite{zomorodian2009computational,edelsbrunner2010computational}. 

\subsection{Basics of Persistent Homology}
\begin{figure}[h]
    \centering
    \includegraphics[width=0.33\textwidth]{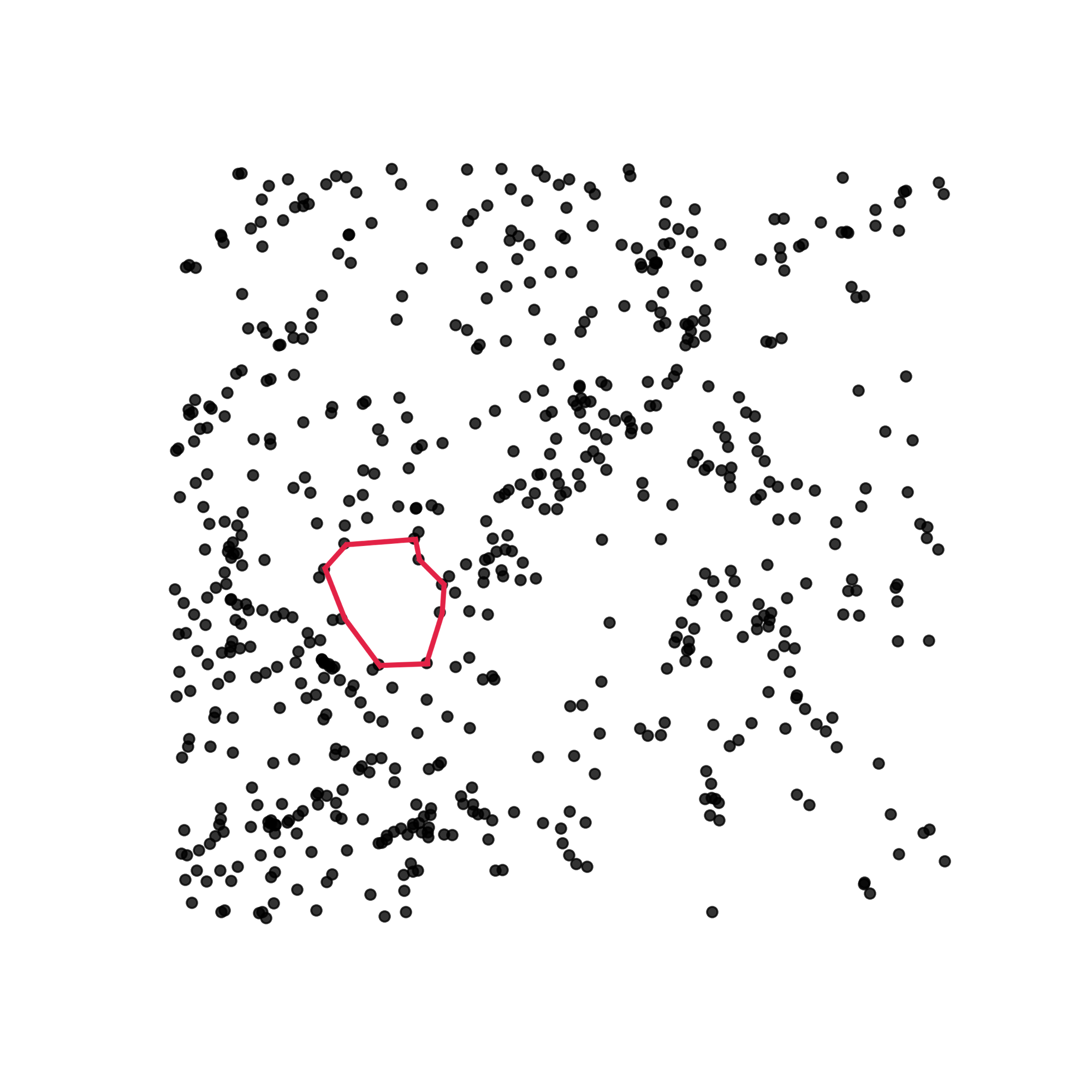}\includegraphics[width=0.33\textwidth]{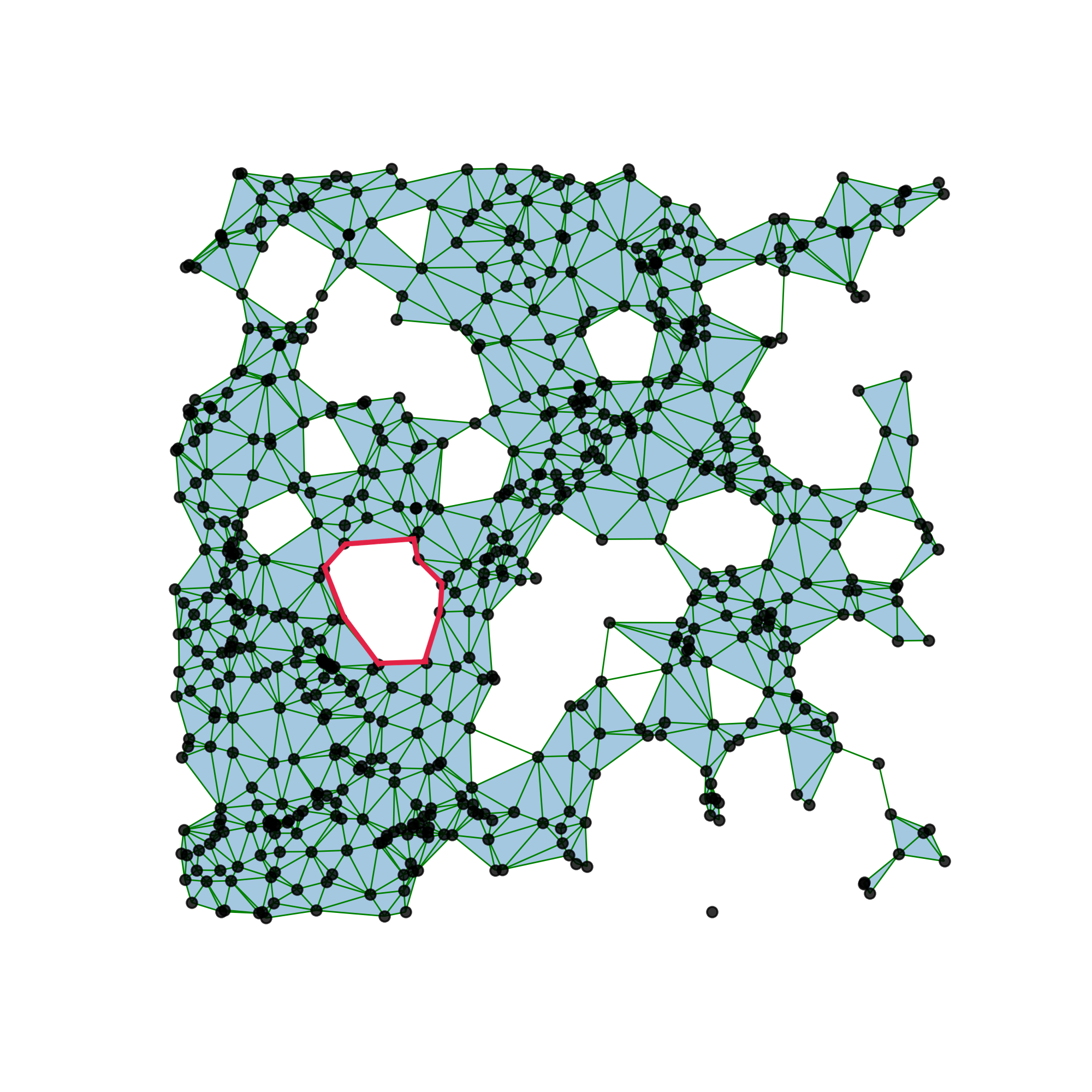}\includegraphics[width=0.33\textwidth]{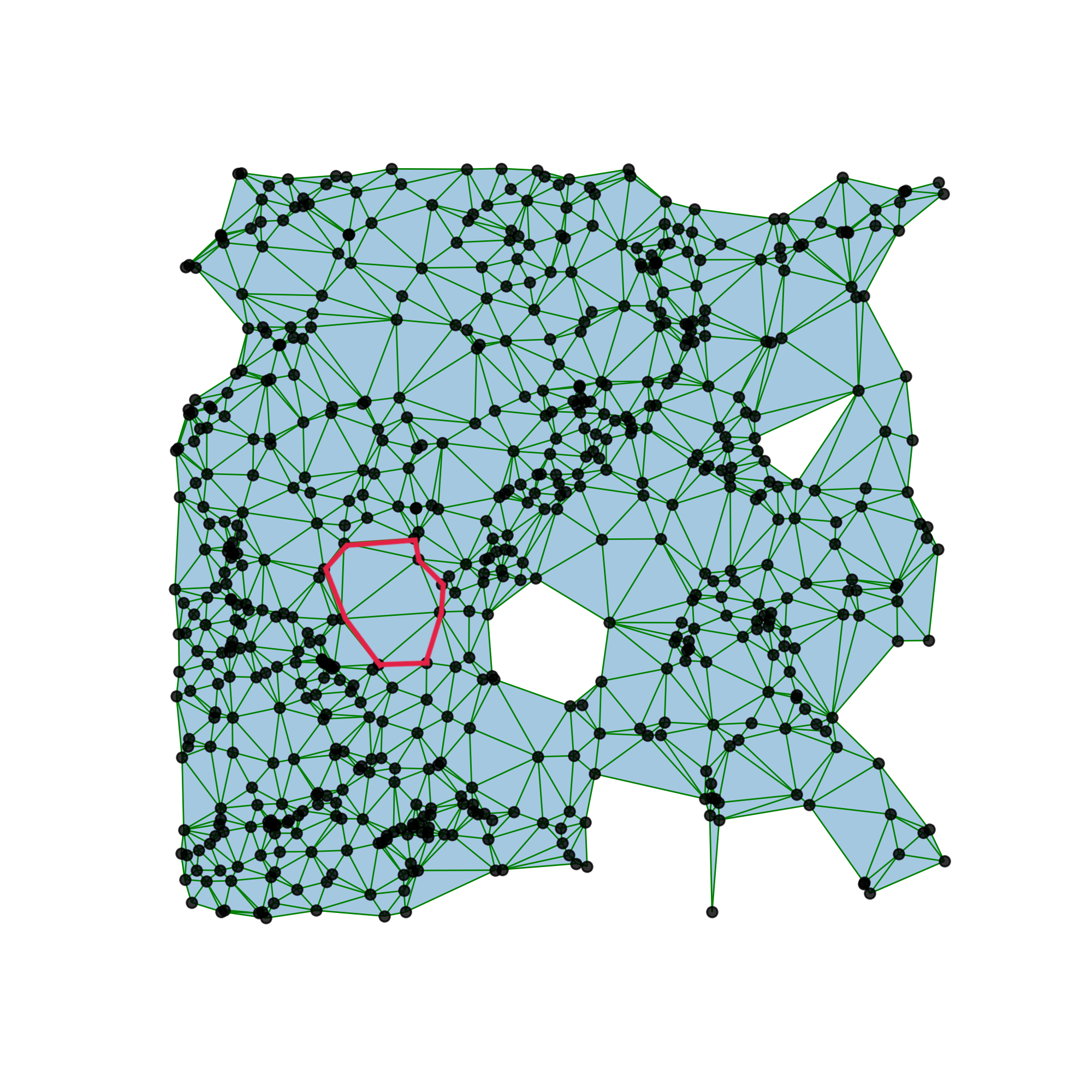}
    \caption{\emph{Left Panel:} projection of dark matter halo positions in a $100\times 100\times 50$ (Mpc/h)$^3$ subbox of one of our simulations. For presentation's sake we consider the 2-dimensional topology (clusters and loops) of this projected data as a function of coarse-graining scale. In red is highlighted a collection of halos that will eventually correspond to a nontrivial homology generator, i.e.\ a hole. \emph{Middle Panel:} at an intermediate coarse-graining scale, most halos have been connected via edges to their nearest neighbors. The red region now corresponds to a nontrivial homology generator. \emph{Right Panel:} at a larger coarse-graining scaling, the ``hole'' in the center diagram has been filled in by triangles. }
    \label{fig:introCartoon}
\end{figure}

The formalism of persistent homology allows us to characterize the multiscale topology of a data set. In our context, the data set under consideration is a catalog of dark matter halos in a cubical box, as a proxy for galaxies as observed by surveys.\footnote{In a companion paper \cite{BiagettiColeYipToAppear}, we exploit a public data set of mock galaxy catalogs built from halo catalogs. The formalism we present here is applied in a very similar way to that data set.}  In other words, we have a ``point cloud'' of halo positions. It is well-known that at late times, the distribution of halos features a rich hierarchy of interlocked voids, filaments, and clusters, often referred to as the cosmic web \cite{Bond:1995yt}. By hierarchy, we refer to the fact that these objects can appear and disappear as a function of scale. Once one has coarse-grained to a scale $L$, voids with radius $R<L$ will be ``filled in.'' Moreover, halos making up the boundary of a void need to connect at scale $L$ in order for the surface of the void to close. Therefore every void can be associated with both a death scale and a birth scale. We illustrate this coarse-graining process in Fig.~\ref{fig:introCartoon}.

Formalizing this intuitive picture and generalizing to lower-dimensional features is possible within the framework of persistent homology.
In the context of this work, the topological features found in three dimensions are clusters, filament loops, and voids. We compute the \emph{distribution} of these features across birth and death scales and use how this distribution reacts to changes in initial conditions to forecast sensitivity to primordial non-Gaussianity.

By construction, the distribution of topological features traces correlation functions of all orders, beyond simply the two-point and three-point functions often considered in cosmology. In order to identify topological features, we embed our data into higher-order structures called simplicial complexes. The mathematical formalism that describes the topology of these complexes is homology.

\paragraph{Simplicial Complexes and Homology.} A simplicial complex is composed of simplices, where a 0-simplex is a vertex, a 1-simplex is an edge between vertices, a 2-simplex is a triangle between three edges, and a 3-simplex is a tetrahedron between four triangles. We refer generally to $p$
-simplices with $0\leq p \leq 3$. A simplicial complex is a collection of simplices that is closed under taking faces of a simplex (e.g.\ for an edge to be present, the two vertices at its ends must be present) and under the intersection of simplices.

Naturally, we associate each halo in a given simulation with a vertex. In order to compute the data's connectivity and higher-order topological aspects, higher-dimensional simplices must also be included. We describe our rules for these simplices in the next section.
Given a simplicial complex, its topology is encoded in the linear operators $\partial_p$ that take a collection of $p$-simplices to its $(p-1)$-dimensional boundary.\footnote{For brevity, we omit the formal definitions of $p$-chains, $p$-cycles, and $p$-boundaries here. For those interested, we define these vectors spaces with coefficients in $\mathbb{Z}_2$.} In words, topological features correspond to collections of $p$-simplices with vanishing boundary that are not themselves the boundary of a collection of $(p+1)$-simplices. An example is shown in the central panel of Fig.\ \ref{fig:introCartoon}. These features can be identified by manipulating the boundary operators $\partial_p$. In particular, one is led to define the \emph{homology groups} $H_p \equiv \ker \partial_p/\textrm{im}\partial_{p+1}$. Informally, each element of $H_p$ is an independent ``hole'' of dimension $p$. Note that a 0-dimensional hole here is a connected component. In three dimensions, one also has filament loops (elements of $H_1$) and voids (elements of $H_2$). Formally, elements of $H_p$ are equivalence classes of collections of $p$-simplices with vanishing boundary.
The ranks of the homology groups, which count the number of independent $p$-dimensional topological features, are called the Betti numbers $b_p \equiv |H_p|$.

\paragraph{Filtrations and Persistent Homology.} As anticipated by our previous discussion, it is natural that the distance between two vertices should be the primary determining factor as to whether the two vertices should be connected by an edge. For example, one way to construct a simplicial complex from a point cloud is to pick a threshold length $L$ and for every pair of vertices separated by distance $d$, connect them with an edge if $d\leq L$. Similar rules may be employed to determine the inclusion of higher-dimensional simplices. It should be immediately clear, however, that the topology of the resulting simplicial complex will be highly unstable with respect to the choice $L$. Moreover, picking a single length scale $L$ at which to view the data set cannot accurately convey a hierarchical distribution of features.

\begin{figure}[h]

     \begin{subfigure}[b]{0.32\textwidth}
         \centering
         \includegraphics[width=\textwidth]{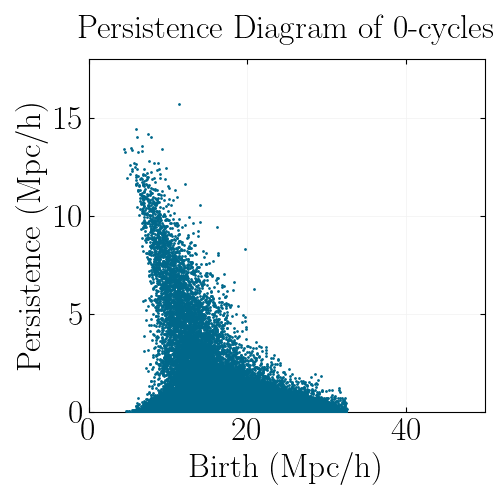}
     \end{subfigure}
     \begin{subfigure}[b]{0.32\textwidth}
         \centering
         \includegraphics[width=\textwidth]{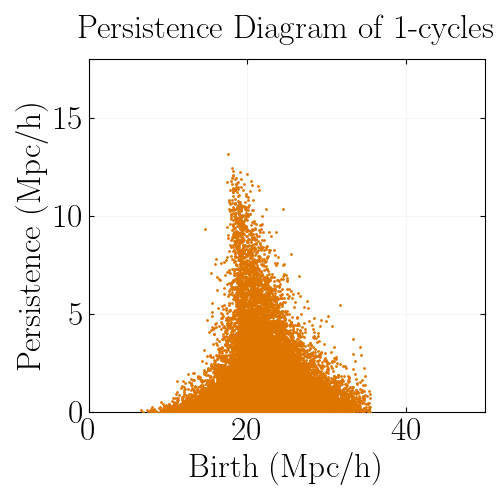}
     \end{subfigure}
     \begin{subfigure}[b]{0.32\textwidth}
         \centering
         \includegraphics[width=\textwidth]{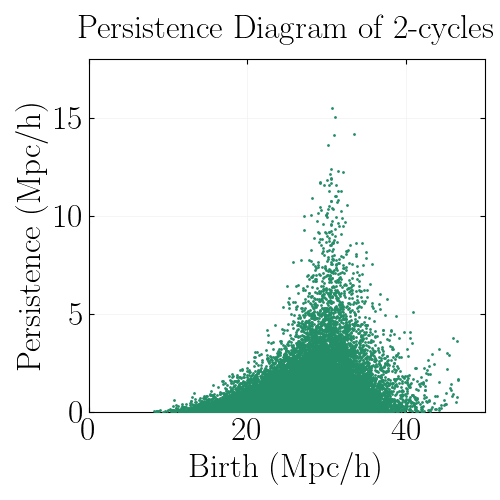}
     \end{subfigure}
     
     \begin{subfigure}[b]{0.32\textwidth}
         \centering
         \includegraphics[width=\textwidth]{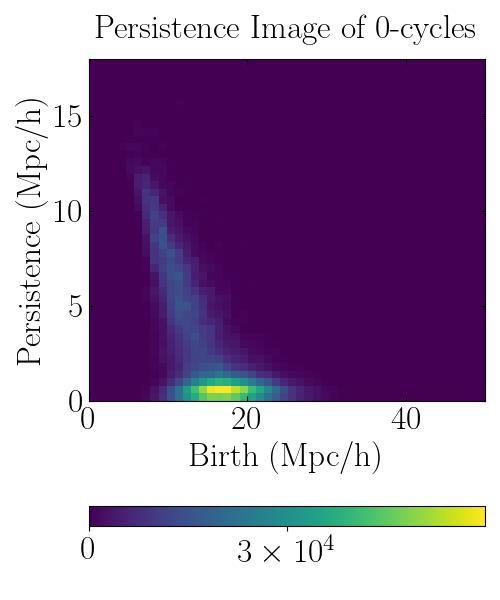}
     \end{subfigure}
     \begin{subfigure}[b]{0.32\textwidth}
         \centering
         \includegraphics[width=\textwidth]{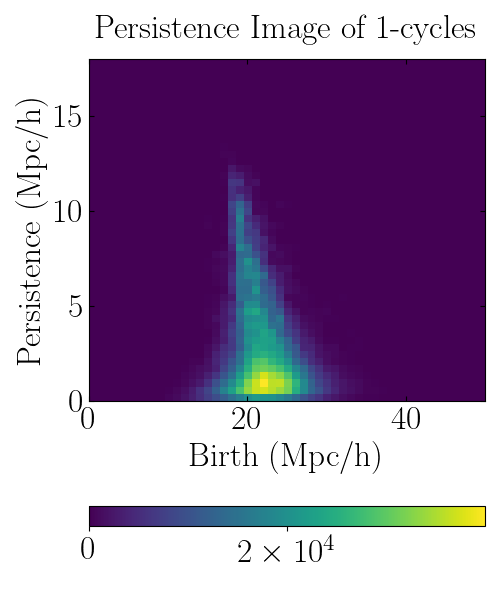}
     \end{subfigure}
     \begin{subfigure}[b]{0.32\textwidth}
         \centering
         \includegraphics[width=\textwidth]{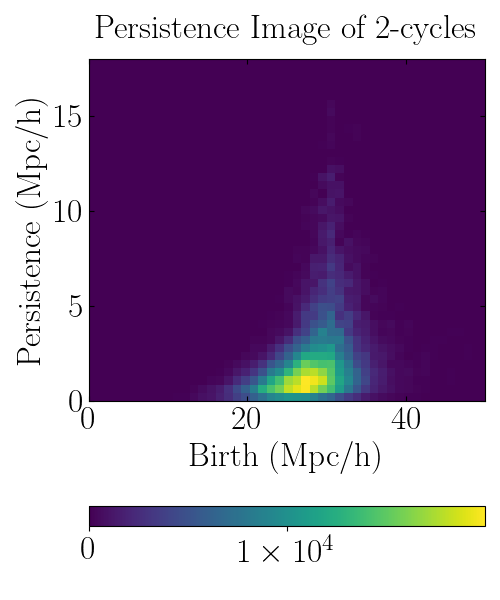}
     \end{subfigure}
          
        \caption{\emph{Top row:} persistence diagram for the $\alpha$DTM$\ell$-filtration for one of our simulations. \emph{Bottom row:} corresponding persistence images, as binned densities with kernel weight proportional to persistence. Weighting the kernel with persistence partially accounts for the fact that the majority of features are short-lived.}
        \label{fig:PD}
        
\end{figure}

These problems are resolved by representing our data set with a \emph{family} of simplicial complexes rather than a single complex. In particular, we use a \emph{filtration}, or a growing family of simplicial complexes. We parameterize a filtration with $\nu$, which to first approximation is a length scale $\nu\sim L$. Each simplex in a filtration is assigned its own value $\nu_0$ at which it is included in the complex. For example, for an edge between two vertices, we might have $\nu_0=d$. Then $\nu$ acts as a filter, so that only simplices with $\nu_0 \leq \nu$ are included at a given point in the filtration. Intuitively, $\nu$ gives the coarse-graining scale at which the data set is considered. As previously mentioned, topological features will be created and destroyed as $\nu$ is increased, with formerly disconnected components joining each other and holes forming and filling in.
Given a filtration, the creation and destruction of \emph{individual topological features} may be computed via matrix reduction methods. This is then \emph{persistent homology} -- we now have access to the fine-grained data of individual homology classes, including their creation, destruction, and length scales in between where the features persist. Note already that this is more refined information than the Betti numbers, which merely count the number of topological features of various dimensions. One way to summarize the topological information in a filtration is via a \emph{persistence diagram}, which is a scatter plot of the filtration parameters $\nu$ at which topological features are created and destroyed. For ease of visualization, we often plot a persistence diagram with the axes $(\nu_{\rm birth}, \nu_{\rm persist})\equiv (\nu_{\rm birth}, \nu_{\rm death}-\nu_{\rm birth})$ since by definition $\nu_{\rm death}> \nu_{\rm birth}$. Examples of persistence diagrams generated for our data are shown in Fig.\ \ref{fig:PD}.

In this work, we use what we call an $\alpha$DTM$\ell$-filtration \cite{chazal2017robust,Biagetti:2020skr}. This refines the filtration described above in two respects. First, rather than allowing an edge between any two vertices, the simplicial structure of a given data set is determined by a Delaunay triangulation. In other words, at a given filtration parameter $\nu$, the complex forms a subcomplex of the Delaunay complex \cite{edelsbrunner1994three}. There is a computational overhead associated with computing the Delaunay complex, but the total number of simplices is dramatically reduced so that the tradeoff is worth it. Our second variation is in the assignment of filtration parameters to simplices. Namely, rather than a purely length-based filtration parameter, we find it useful to include information related to the local density of points in our data. In particular, this affords our filtration more robustness against outliers.

\begin{figure}
    \centering
    \includegraphics[width=0.5\textwidth]{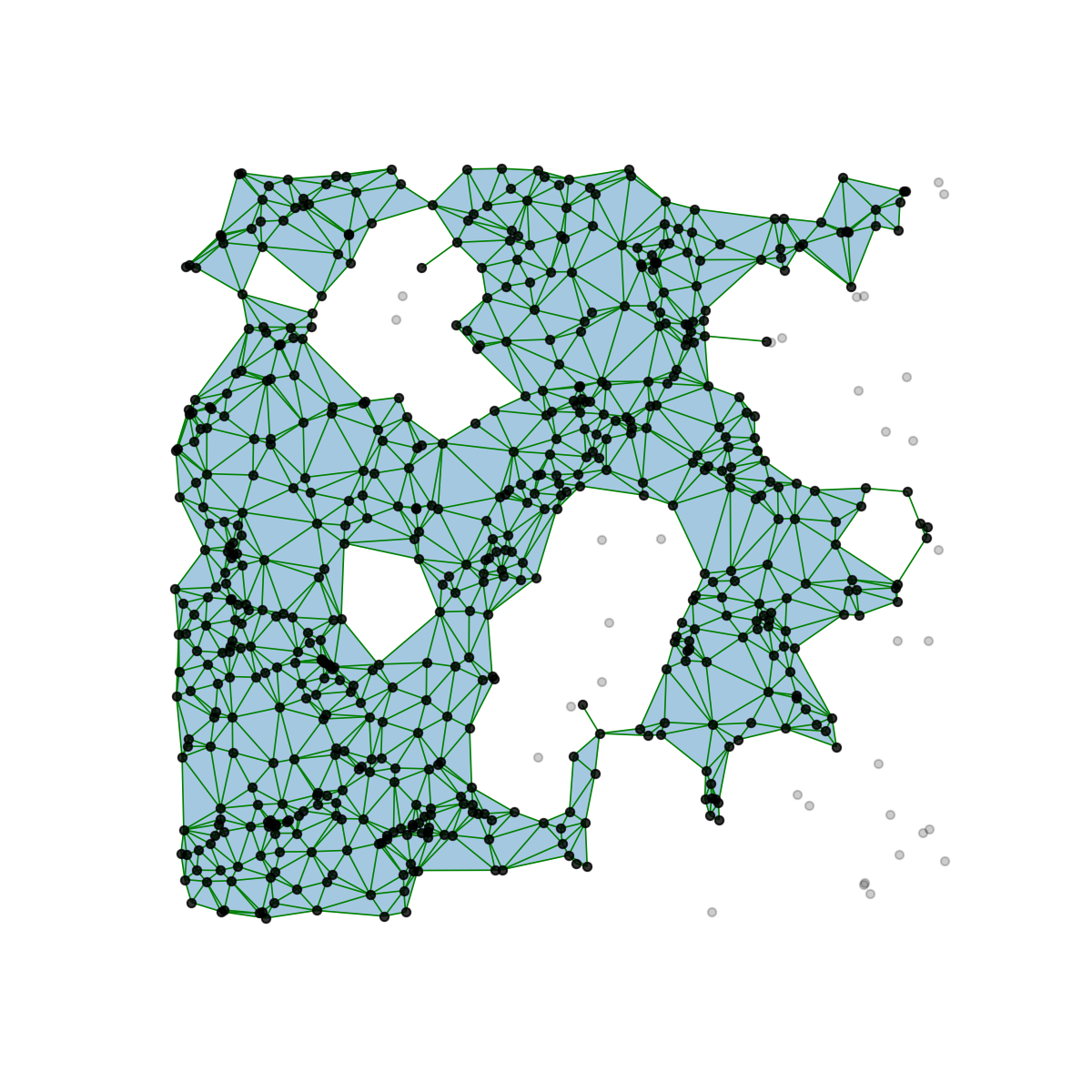}
    \caption{In the $\alpha$DTM$\ell$-filtration, outliers are delayed in creating simplices. This leads to larger-scale features compared to an $\alpha$-filtration. Here we shown the same point distribution as in Fig.\ \ref{fig:introCartoon}, but with an $\alpha$DTM$\ell$-filtration. Points that still have not been added to the complex because their DTM values are too large are shown in gray. }
    \label{fig:cartoonDTM}
\end{figure}

To accomplish this, we use the ``distance-to-measure'' or DTM function
\begin{equation}
\label{eqn:DTM}
    {\rm DTM}(x)=\left(\frac{1}{k}\sum_{y\in N_k(x)}\left|\left| x-y \right|\right|^p\right)^{1/p}
\end{equation}
where $k>0, p>0$, $N_k(x)$ is the set of the $k$ nearest neighbors to $x$ in the point cloud. The interpretation of DTM$(x)$ is that it quantifies the extent to which point $x$ is an outlier within the point cloud -- DTM$(x)$ is large with its $k$-nearest neighbors are far away, and it is small when its $k$-nearest neighbors are close by. A given vertex at position $x$ is assigned $\nu={\rm DTM}(x)$ as its filtration value. For edges, we mix the DTM value with edge length, the interpretation being that a point $x$ is surrounded by a ball of radius
\begin{equation}
    \label{eqn:lDTM-mix}
    r_x(\nu)= 
    \begin{cases}
    \left(\nu^q - {\rm DTM}(x)^q\right)^{1/q} ,& \text{if } \nu\geq {\rm DTM(x)}\\
    0,              & \text{otherwise}
\end{cases}
\end{equation}
with $q>0$, and an edge is included when the relevant balls overlap. Crucially, the edges are still taken from those present in the Delaunay complex. Higher-order simplices are then added to the filtration when all necessary faces are present. The effect of DTM$(x)$ on $r_x(\nu)$ is that the growth of balls around outliers is impeded. For example, if there are sparse points in the interior of a void, using $r_x(\nu)$ from eqn.\ \ref{eqn:lDTM-mix} rather than say $r_x(\nu)=\nu$ ensures that the void is not prematurely filled by simplices associated to outlier points. Therefore we track larger-scale features than in an $\alpha$-filtration, see Fig.\ \ref{fig:cartoonDTM} for an illustration.

In this work, we take $k=15$, $p=q=2$, as in \cite{Biagetti:2020skr}. Currently these choices are based on heuristics. For example, we observe that for our data the DTM distribution with $k=15$ is approximately normal. We note in passing that $p,q$ can theoretically be tuned via gradient-based methods, which we leave to future work.

Additionally, we note that the DTM function also gives us precise information regarding the distances to nearest neighbors of a point. Recently, the distribution of such distances has been considered as a summary statistic in its own right \cite{Banerjee:2020umh,Banerjee:2021cmi,Banerjee:2021hkg}. Results from those works suggest that it could be beneficial to simultaneously consider the topology for different values of $k$.

\subsection{From Persistence Diagrams to Summary Statistics}

As mentioned, the topological information content of a filtration is summarized via a persistence diagram. As a collection of points, a persistence diagram is not directly amenable to statistical analysis. We would prefer a topological summary that lives in a vector space. There are many proposals for how to perform this map (or more generally how to compute an inner product on the space of persistence diagrams) \cite{carriere2015stable,bubenik2015statistical,adams2017persistence,kalivsnik2019tropical} -- in general, any permutation-invariant function is allowed, and the map can be parameterized by a neural network \cite{zaheer2017deep,carriere2020perslay}. Persistence images \cite{adams2017persistence}, which are essentially smoothed histograms of a persistence diagram, are particularly useful for visualizing the density of topological features. In the smoothing kernel, one usually includes a persistence-dependent weight such as $\nu_{\rm persist}$ or $\log(\nu_{\rm persist})$ to emphasize long-lived features, or alternatively because most features are short-lived. Some examples of persistence images are shown in Fig. \ref{fig:PD}.
In our present context, we lack the data volume to train a neural network, and favor a compressed representation so that covariance matrices may be estimated. We therefore use as summary statistics histograms representing the distribution of births and distribution of deaths for cycles of various dimension. We employ cutoffs at the 99.9th percentile for a fiducial simulation,
to remove sparse outliers. Subsequently, each distribution is summarized by $5$ bins, for a total of $30$ bins. This is to be compared with $120$ simulations of volume $1~\left({\rm Gpc/h}\right)^3$. We expect that more information can be extracted from an optimized vectorization of persistence diagrams. 
Optimizing these representations will require a larger amount of data than we consider in the present work.

\section{Primordial non-Gaussianity}\label{sec:phhalo}

Primordial non-Gaussianities can be sourced by a wide variety of interactions taking place during inflation. 
The leading non-Gaussianity is the three-point correlation function of the primordial curvature perturbation. Its information is encoded in the bispectrum in Fourier space. Although the exact form of the bispectrum is model-dependent, non-Gaussianities can be classified by the triangle configuration where their contribution is strongest. The most popular primordial non-Gaussianities are related to so-called local and equilateral shapes, which peak on squeezed and equilateral triangle configurations, respectively. Separable templates for these shapes have been introduced \cite{Babich:2004gb} for a fast implementation in observations. Here we define these templates and give a brief review of how they are implemented in the numerical simulations that we use for our study.

\paragraph{Local type.} Primordial non-Gaussianity of the local type can be written as a Taylor expansion around a Gaussian field of the primordial fluctuations \cite{Salopek:1990jq}, around a given position $\bx$:

\begin{equation}\label{eq:locpng}
    \zeta(\bx) = \zeta_{\rm G}(\bx) + \frac 35\fnlloc\left(\zeta^2_{\rm G}(\bx) - \langle \zeta_{\rm G}^2\rangle \right) +\mathcal O\left(\zeta_{\rm G}^3\right) ,
\end{equation}
with $\zeta$ the comoving curvature perturbation, $\zeta_{\rm G}$ a Gaussian random field, and $\fnlloc$ parametrizing the amplitude of non-Gaussianity.\footnote{A set of consistency relations constrain this type of non-Gaussianity to be very small in amplitude for single-field models of inflation \cite{Maldacena:2002vr,Creminelli:2004yq,Creminelli:2012ed}, while it would be detectable in case more than one field is responsible for generating $\zeta$  (see \cite{Bartolo:2004if} for a review). Experimental upper bounds on $\fnlloc$ (both from CMB \cite{Planck:2019kim} and LSS \cite{Castorina:2019wmr}) are of the order of $|\fnlloc| \lesssim 10$ at $95\%$ confidence level.} The expansion Eq.\eqref{eq:locpng} generates the following bispectrum
\begin{equation}
     B_\zeta^{\rm local}(k_1,k_2,k_3)  = \frac{6}{5} \fnlloc \biggr[P_\zeta(k_1)P_\zeta(k_2)+2\,\,\mbox{perms.}\biggr],
\end{equation}
which peaks on squeezed triangle configurations of the bispectrum, i.e. when correlating a long-wavelength mode with short-wavelength modes. This parameterizes a physical correlation between small-scale physics and the large-scale gravitational potential, induced during inflation.  Such a correlation is known to generate a scale-dependent enhancement or suppression (depending on the sign of $\fnlloc$) at very large scales in the galaxy two-point correlation function (see \cite{Biagetti:2019bnp} for a review), which allows constraining this type of non-Gaussianity quite well using galaxy surveys. 

\paragraph{Equilateral type.} Another type of primordial non-Gaussianity is so-called \emph{equilateral}, which is characterized by a bispectrum that peaks in the equilateral configuration, i.e.\ when all sides of the triangle are equal. A popular model producing this bispectrum is the Dirac-Born-Infeld (DBI) model \cite{Silverstein:2003hf,Alishahiha:2004eh, Chen:2006nt}. We parameterize this type of non-Gaussianity with the usual template \cite{Creminelli:2005hu}
\begin{align}\label{eq:eqpng}
 B_\zeta^{\rm equi}(k_1,k_2,k_3) & = \frac{18}{5} \fnleq\left[-P_\zeta(k_1)P_\zeta(k_2)-2\,\,\mbox{perms.} - 2P_\zeta^{2/3}(k_1)P_\zeta^{2/3}(k_2)P_\zeta^{2/3}(k_3)+
   \right.\nonumber\\
    & \left. P_\zeta^{1/3}(k_1)P_\zeta^{2/3}(k_2)P_\zeta(k_3)+5\,\,\mbox{perms.}\right].
\end{align}
Here the normalization of the amplitude $f_{\rm NL}$ is fixed such that $\fnleq \equiv \fnlloc$ in the equilateral limit.

\subsection{Implementation in N-body simulations}

N-body simulations of the universe on large scales solve the  non-linear equations describing gravitational evolution from an early time, where perturbations are linear, until late times. Initial displacements of the particles can be set using  Lagrangian perturbation theory (LPT) at an early redshift $z \sim 100$. If the initial conditions of the simulation are Gaussian, particles are randomly distributed on a mesh grid and then displaced by the LPT displacement field $\psi$ \cite{Bernardeau:2001qr}. Implementing non-Gaussian initial conditions therefore requires an additional displacement by $\propto f_{\rm NL}\nabla \zeta_{\rm G}$ on the particles as they are assigned to the grid, before being displaced by $\psi$.

In the case of local type primordial non-Gaussianity, such displacement is trivial to implement, as Eq. \eqref{eq:locpng} is defined for each position $\bx$, so it can be applied particle-by-particle by just converting $\zeta$ to the gravitational potential $\Phi$. Note that primordial correlations of $4$ and more points are generated by this procedure. For example, the quadratic term in Eq.~\eqref{eq:locpng} generates a primordial 4-point function of the form $\langle\zeta\zeta\zeta\zeta\rangle \sim (\fnlloc)^2\langle\zeta_G \zeta_G\rangle\langle\zeta_G\zeta_G\rangle$. This is somewhat artificial in the sense that we neglected all higher powers of $\zeta_G$ in equation~\eqref{eq:locpng} (such as $g_{\text{NL}}\zeta_G^3$). These ignored terms could generate higher-order primordial correlation functions of the same order or larger than those generated by the terms we kept. On the other hand, for small values of $\fnlloc$, higher-order primordial correlation functions of $\zeta$ are suppressed by powers of $\fnlloc\zeta$, though they could in principle have a small effect at small scales. We cannot  rule out the possibility that the effects that we observe are in part due to these higher-order correlation functions. However, they are all sourced by the direct coupling between long and short modes introduced in Eq.~\eqref{eq:locpng}, so we are constraining the effect of such a coupling.

The equilateral case is slightly more complicated, but has been extensively studied in the past (see \cite{Scoccimarro:2011pz} and \cite{Angulo:2021kes} for reviews).  In this case, the generation of
initial conditions involves finding a random field that satisfies the bispectrum shape of Eq. \eqref{eq:eqpng}, which is not a unique operation.  This can be done by introducing an appropriate quadratic displacement. However, this procedure also sources small higher-order primordial correlation functions. A generic algorithm was developed in \cite{Scoccimarro:2011pz} (see also \cite{Fergusson:2010ia}), in which bispectrum template is decomposed into factorizable functions. This allows one to generate a one-parameter solution for the inversion. The parameter can be used to constrain the appropriate behaviour of the equilateral shape in the squeezed limit.

Primordial correlations at all higher orders are generated by this algorithm. These are somewhat more artificial than the case of local non-Gaussianity, since they are simply a byproduct of the procedure used to compute the initial displacements. On the other hand, they are also suppressed by powers of $f_{\rm NL}\zeta$. Again, the effects we observe could be due in small part due to these primordial higher-order correlations. However, the coupling between scales in this case is very different from the ``local'' one: initial conditions are adiabatic and the coupling between long $k_L$ and short $k_s$ wavelength modes is suppressed by $k_L^2/k_s^2$ \cite{Maldacena:2002vr,Creminelli:2004yq,Creminelli:2012ed,Peloso:2013zw,Kehagias:2013yd,Creminelli:2013mca}. Thus, we are studying the effect of non-Gaussianity in the absence of such a coupling.

\section{Methodology}\label{sec:implementation}

In this section, we review all steps of the method, from defining these summary statistics, the dataset that we use and the Fisher formalism we employ, along with several consistency tests.

\subsection{Dataset and persistence calculation}

\begin{figure}[h]

     \begin{subfigure}[b]{0.32\textwidth}
         \centering
         \includegraphics[width=\textwidth]{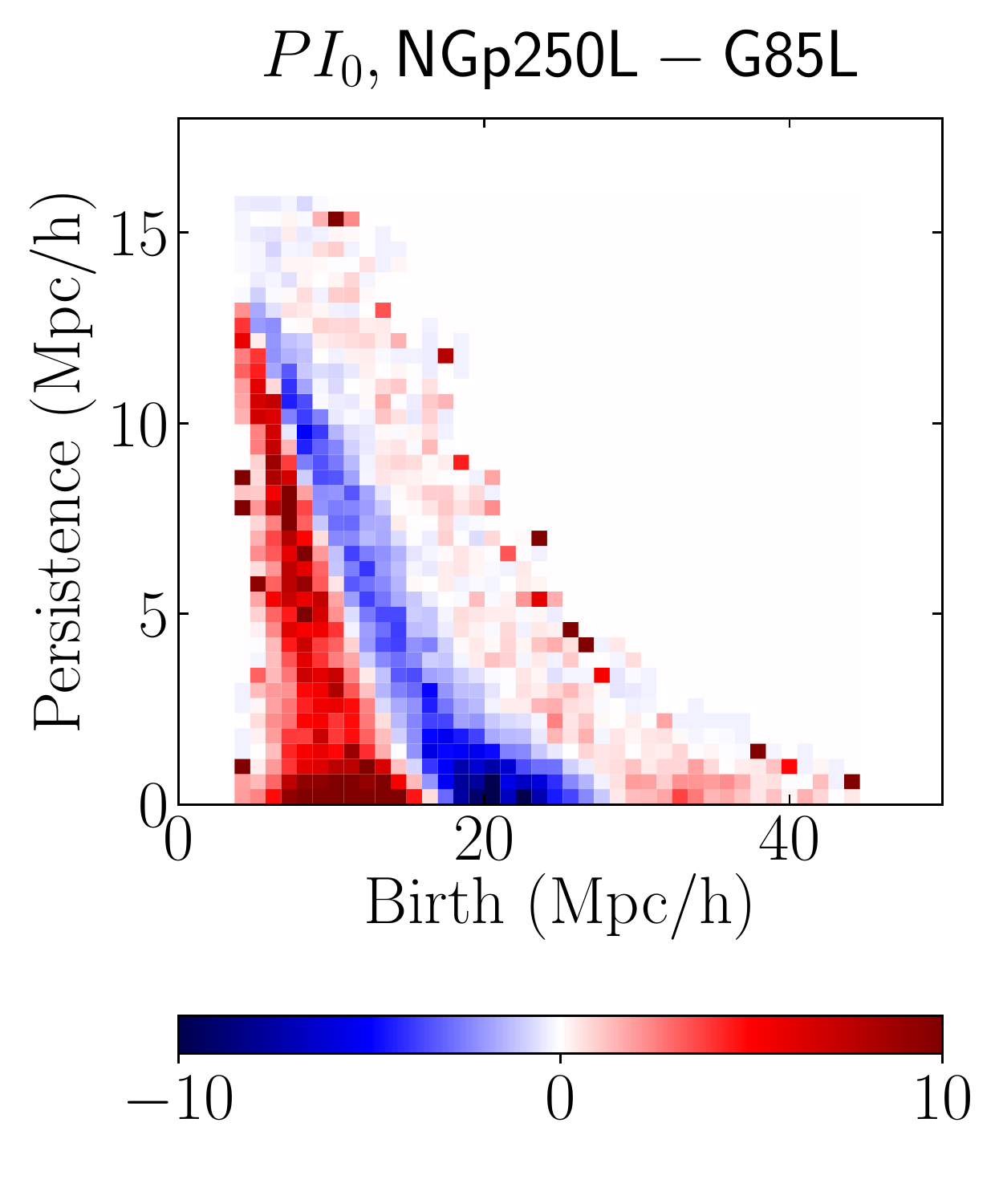}
     \end{subfigure}
     \begin{subfigure}[b]{0.32\textwidth}
         \centering
         \includegraphics[width=\textwidth]{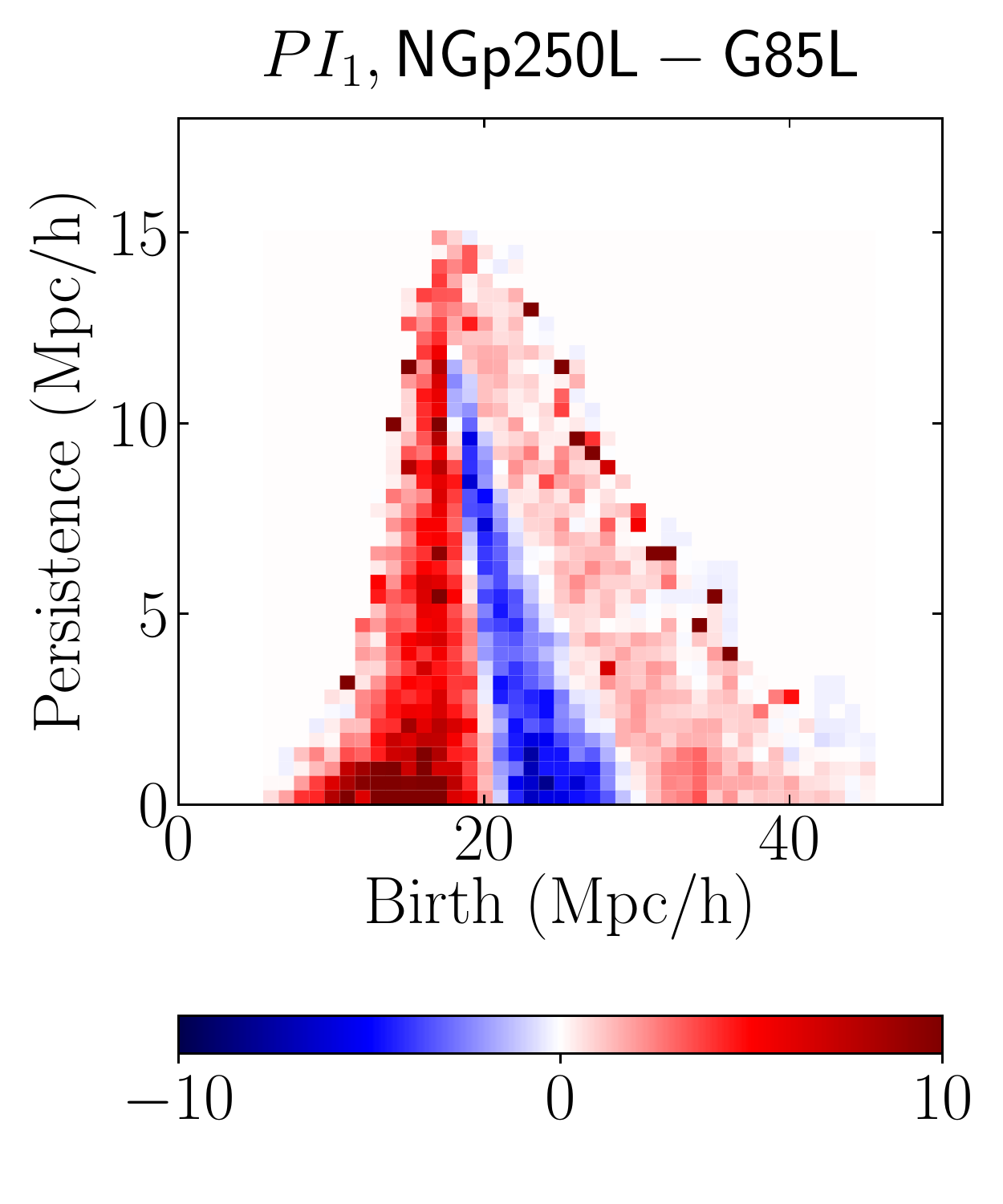}
     \end{subfigure}
     \begin{subfigure}[b]{0.32\textwidth}
         \centering
         \includegraphics[width=\textwidth]{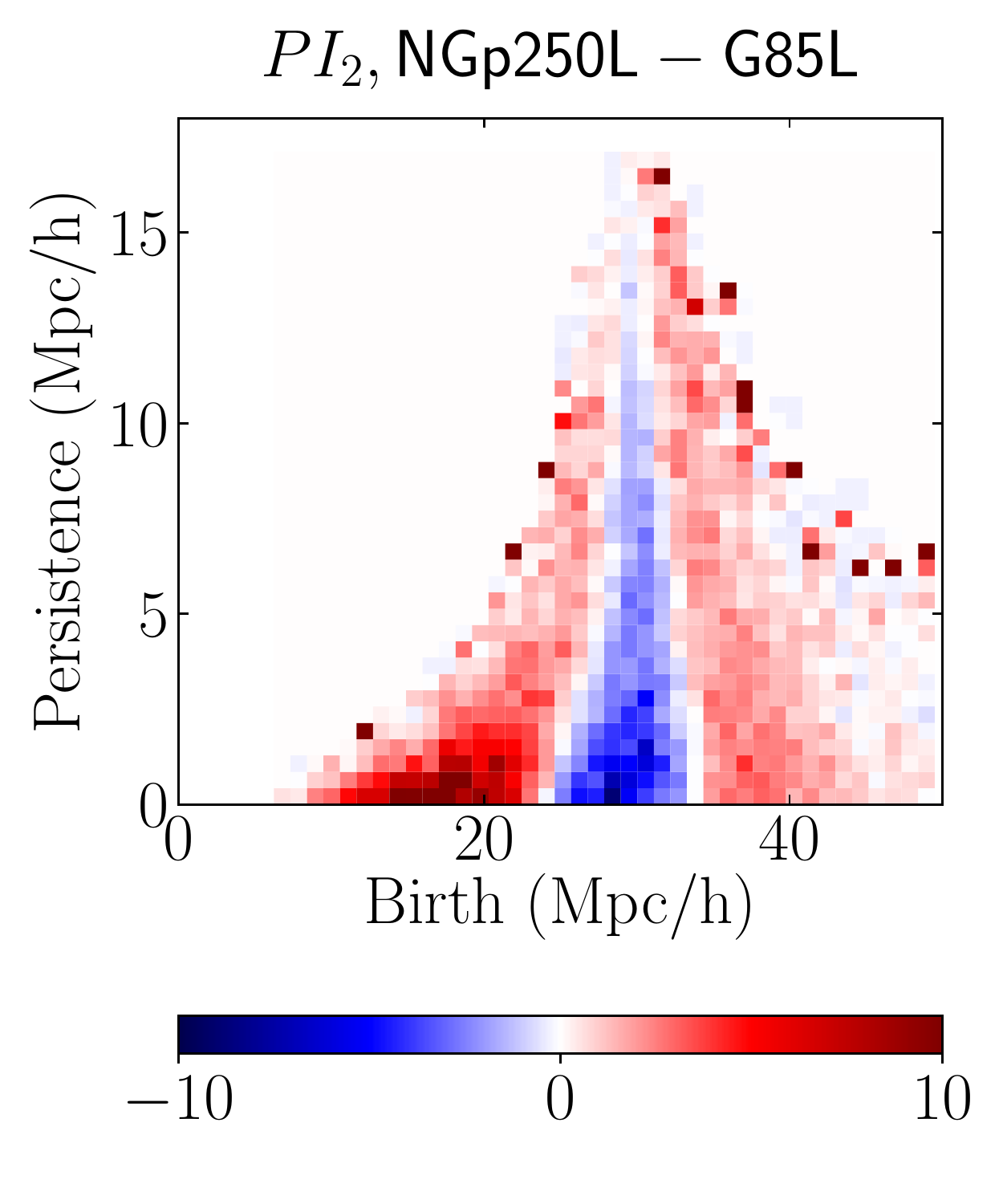}
     \end{subfigure}
     
     \begin{subfigure}[b]{0.32\textwidth}
         \centering
         \includegraphics[width=\textwidth]{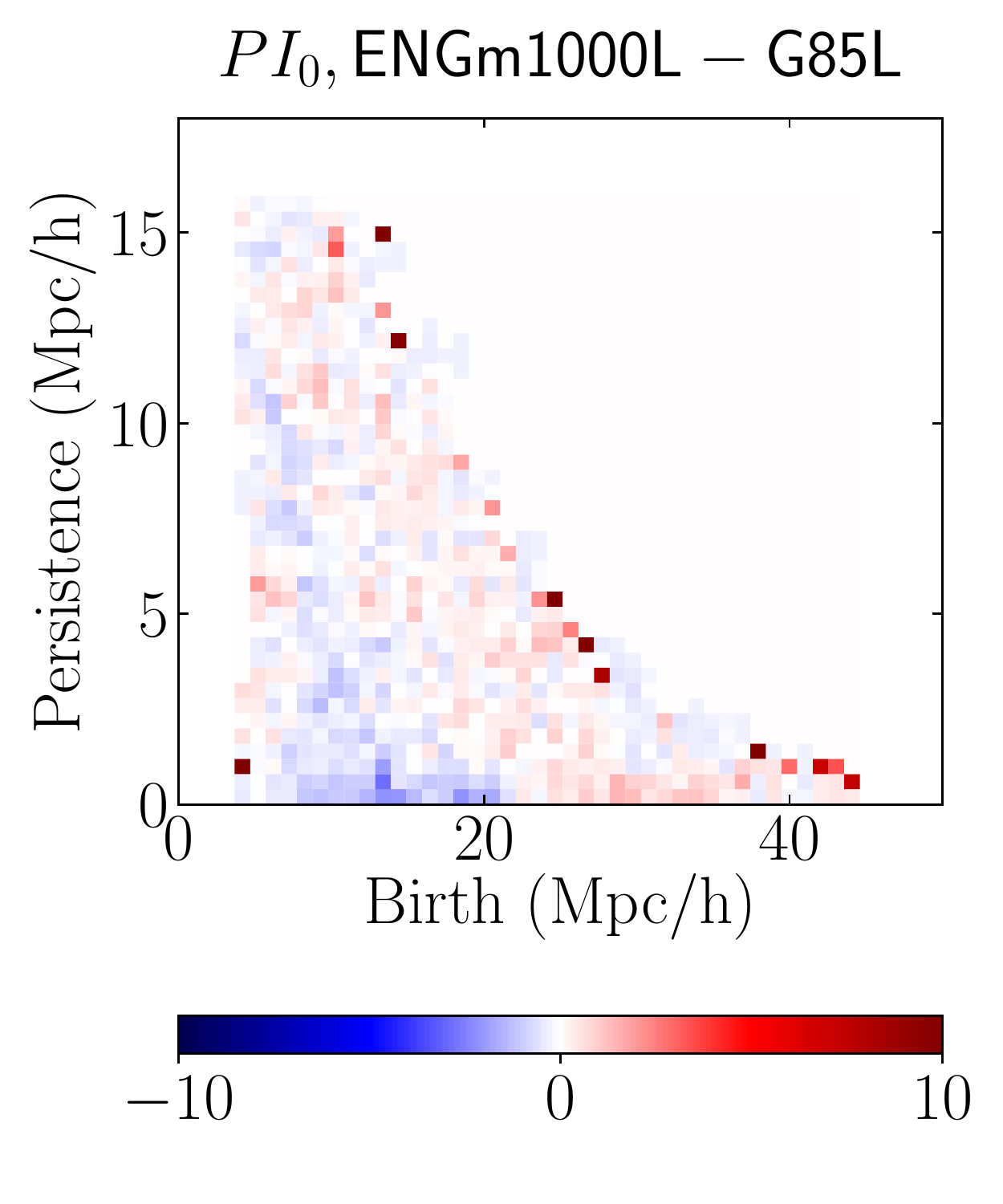}
     \end{subfigure}
     \begin{subfigure}[b]{0.32\textwidth}
         \centering
         \includegraphics[width=\textwidth]{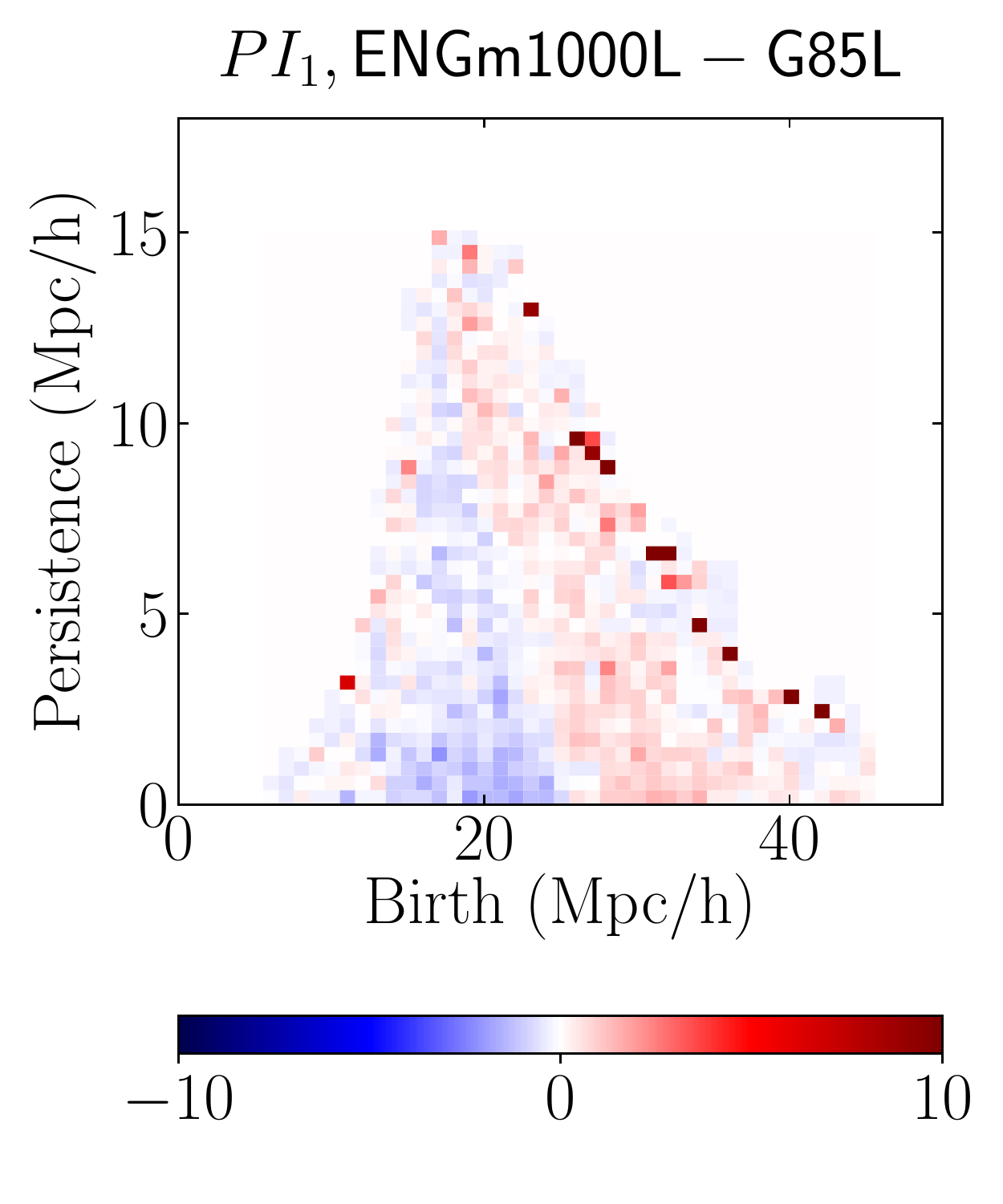}
     \end{subfigure}
     \begin{subfigure}[b]{0.32\textwidth}
         \centering
         \includegraphics[width=\textwidth]{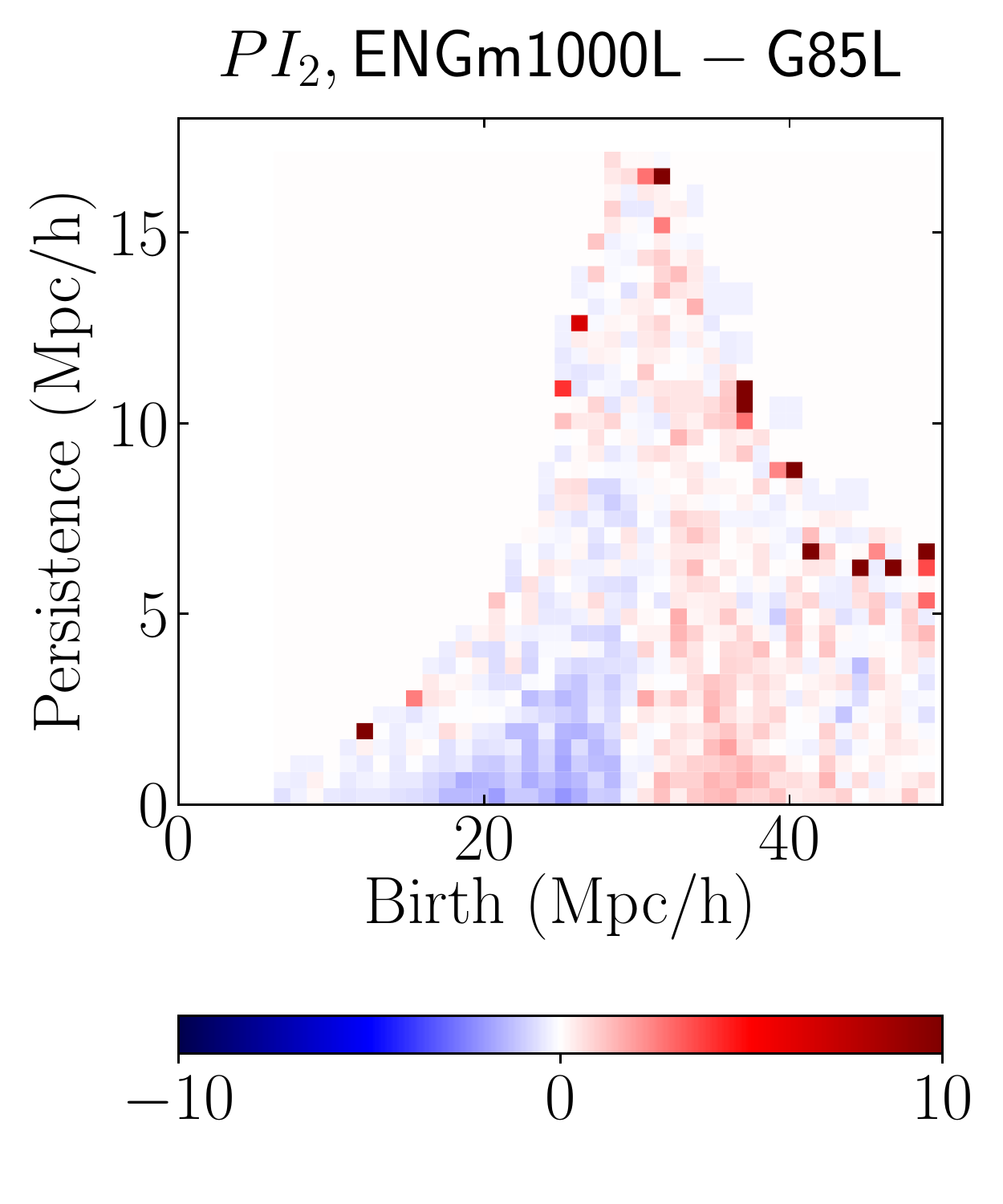}
     \end{subfigure}

    \caption{Difference between the persistence image of the local non-Gaussianity \textsf{NGp250L} and the Gaussian persistance image \textsf{G85L} normalized by the standard deviation of the G85L pixels for each point, at redshift $z=1$ using the $\alpha$DTM$\ell$-filtration. Here the subtraction is done over the average of each full data set.}
    \label{fig:PIloc}
\end{figure}

To perform our analysis we use the \eos \dataset,\footnote{Information on the \eos suite is available in  \cite{Biagetti:2016ywx} and at \texttt{https://mbiagetti.gitlab.io/cosmos/nbody/eos/}.}  which includes simulations with primordial non-Gaussianity of equilateral and local type. The initial particle displacement is generated using the public parallel code \texttt{L-PICOLA}  \cite{howlett2015picola} for realizations with equilateral non-Gaussianity and with \texttt{2LPTic} \cite{Scoccimarro:1997gr,Crocce:2006ve} for realizations with Gaussian and local non-Gaussian initial conditions. The cosmology is flat $\Lambda$CDM with $\sigma_8=0.85$, $h=0.7$ and $\Omega_m=0.3$. Initial condition are generated at $z_{in}=99$. The public code \texttt{Gadget2} \cite{Springel:2005mi} is used to evolve $1536^3$ particles in a cubic box of $2$ Gpc$/h$ per side. Each box has $15$ different realizations. Gaussian initial conditions simulations are named \textsf{G85L} hereafter. Simulations with non-Gaussian initial conditions of local and equilateral type are initialized with $\fnlloc=10$ and $\fnleq=-30$ respectively and we refer to them as \textsf{NG10L} and \textsf{ENGm30L}. For visualization purposes (see Figure \ref{fig:PIloc}) we also use simulations with stronger signal with $\fnlloc=250$ and $\fnleq=-1000$, labeled as \textsf{NGp250L} and \textsf{ENGm1000L}, respectively.
For our analysis, we identify halos in each simulation using the code \texttt{Rockstar} \cite{Behroozi:2011ju}, identifying candidate halos with a minimum of $50$ particles using a Friends-of-Friends (FoF) algorithm with a
linking length $\lambda = 0.28$ at redshift $z=1$.  This results in halos with minimum mass $M_{\rm min}=9.2 \times 10^{12} M_\odot/h$, which would host a galaxy population roughly compatible to the BOSS CMASS sample (see \cite{Grieb:2015bia} for an application of an HOD model to a similar halo population). As explained just below, we divide each simulation box into $8$ sub-boxes of $1 ($Gpc$/h)^3$ volume, which makes a total of $120$ sub-boxes per cosmology. 

\paragraph{Implementation of redshift errors.} Part of our analysis is devoted to halos in redshift space. We are going to adopt two different prescriptions to displace halos from real to redshift space: 
in the plane-parallel approximation we displace halos along one of the axes parallel to a side of the box $\hat{x}$, $\hat{y}$, or $\hat{z}$. In the other case, we displace them along the vector between the observer (at the origin of the box) and the halo $\hat{r}$.

We also mimic redshift errors by trading redshift uncertainties with velocity uncertainties at fixed redshift in the box. We distinguish two hypothetical cases, corresponding to a spectroscopic sample and a photometric sample. Since the typical velocity dispersion of the \eos halos is around $\sigma_v\sim300$ km/s, we assume a spectroscopic survey which has an error on redshift estimation corresponding to an uncertainty on the velocity of about $5$ times smaller than $\sigma_v$. Hence, we displace halos using a random Gaussian distribution centered on the redshift space position and with standard deviation $\sigma_{\rm spec} \sim \sigma_v/5 = 60$ km/s. As for the photometric sample, we take an hypothetical error on redshift determination of $\Delta z /(1+z) = 0.01$, which translates to $\sigma_{\rm photo} = 3000$ km/s. 
In this case, we do not displace halos by their peculiar velocities since they are negligible with respect to $\sigma_{\rm photo}$. In summary, we have the following structure for the plane parallel approximation case:
\begin{align}\label{eqn:zerrors1}
    \bs &= \bx + \frac{v_z}{\mathcal H}\,\hat{\mathbf{n}} + \frac{v_{\rm spec}}{\mathcal H}\, \hat{\mathbf{n}} \qquad\qquad\mbox{Spectroscopic Error},\\
    \mbox{}&\nonumber\\
    \bs &= \bx  + \frac{v_{\rm photo}}{\mathcal H}\, \hat{\mathbf{n}} \qquad\qquad\qquad\,\,\,\,\mbox{Photometric Error},\label{eqn:zerrors2}
\end{align}
where $\bs$ is the position in redshift space, $\mathcal H$ is the Hubble parameter, $\hat n$ is the line of sight, corresponding to the z-axis for the plane parallel approximation and to central vector for the wide angle observer. The error is represented by $v_X$ with $X=\{{\rm spec},{\rm photo}\}$, which is a random variable drawn from a normal random distribution with variance $\sigma_X$ and zero mean. 
\paragraph{Persistence calculations.}
We construct the $\alpha$DTM$\ell$-filtration using the public code\footnote{\texttt{https://gitlab.com/mbiagetti/persistent\_homology\_lss}} developed in \cite{Biagetti:2020skr}. The algorithm is run on sub-boxes of $1$ Gpc$/h$ per side of each simulation box, for a total of $8$ sub-boxes per simulation. As discussed in \cite{Biagetti:2020skr}, most of the features are born and die at scales much smaller than the size of each sub-box, hence effects from the boundary of a given sub-box may be neglected.\footnote{The cutoff at the $99.9$th percentile further removes sparse effects.} Interestingly, 
the signature of primordial non-Gaussianity of the local type does not peak on large scales as in the case of the galaxy/halo two point correlation function, but around a birth scale of order $\mathcal O(10)$ Mpc/h. We show the difference in persistence images between $\fnlloc=250$ and $\fnlloc=0$ simulations in Figure \ref{fig:PIloc}. The strongest differences are seen between birth scales of around $\sim10$ and $\sim 30$ Mpc/h. As discussed in \cite{Biagetti:2019bnp}, the observed shift to smaller birth scales might be connected to the change in inter-halo distance caused by primordial non-Gaussianity and for $2$-cycles it has some similarities to the change expected in the void size distribution found in \cite{Kamionkowski:2008sr}. We observe a smaller change for equilateral non-Gaussianity (bottom panels) and a reversed behaviour: features are born at larger scales in the presence of primordial non-Gaussianity. This might simply depend on the different sign chosen for $\fnleq$. The difference in amplitude between equilateral and local shapes might be due to the effect of scale dependent bias, but we deserve a more accurate investigation to future work.  In summary, we produce  persistence diagrams for $0$-, $1$- and $2$-cycles for the $\alpha$DTM$\ell$-filtration for each of the $8$ sub-boxes, for each of the $15$ realizations, so as to have effectively $120$ realizations in total per cosmology.

\subsection{Summary Statistics}
Previously, \cite{Biagetti:2020skr} extracted information from persistence diagrams via the Betti numbers and several empirical distribution functions.

These statistics contain explicit information about how the feature scales of birth, death, and persistence respond to primordial non-Gaussianity. Given 3 dimensions of homology and 3 types of empirical distribution functions, the number of bins in these statistics was quite large, presenting an obstruction to building a covariance or Fisher matrix. Tests built on hand-crafted templates nevertheless proved successful in identifying small levels of primordial local non-Gaussianity.
As introduced in Section \ref{sec:PHreview}, we construct a low-dimensional summary statistic via histograms of birth and death scales of cycles of each dimension.\footnote{Note that for such an operation there are only two independent histograms that can be built out of birth, death and persistent scales. We choose to use birth and death as they are the direct product of the computation and because from the persistence images Figure 
\ref{fig:PD} it is clear that non-Gaussian signatures are most evident for varying birth and death scales.} 
For varying cosmology, we expect to have measurable differences between $\sim10$ and $\sim 40$ Mpc$/h$, given what is observed in the persistence image (cfr. Figure~\ref{fig:PIloc}). 
\begin{figure}
    \centering
    \includegraphics[width=1.\textwidth]{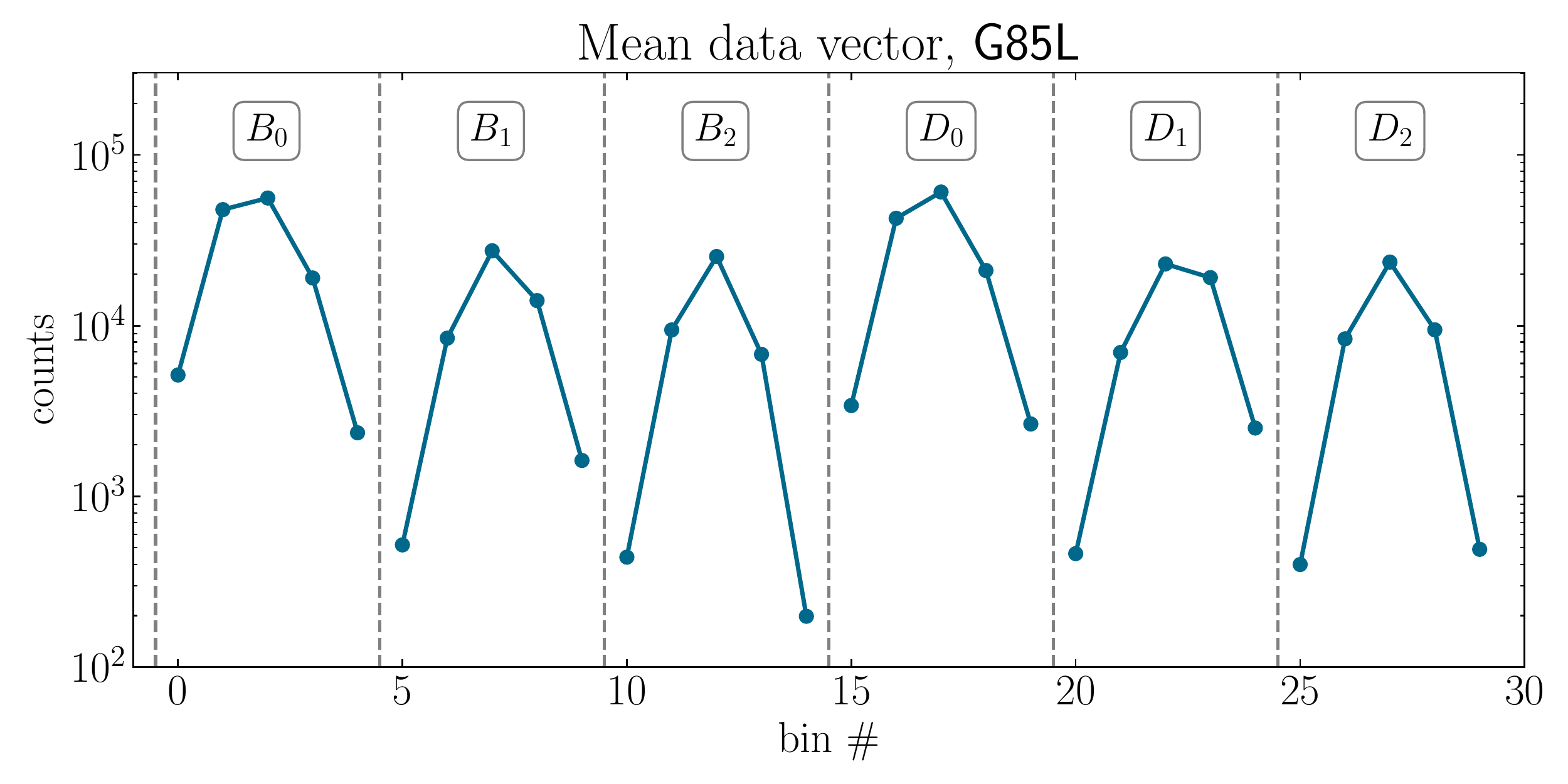}
    \caption{The mean data vector for our fiducial cosmology, \textsf{G85L}. The statistic corresponds to coarse histograms of births and deaths of $0$-, 1-, and 2-homology classes. At this level of binning, each bin contains at least $\mathcal{O}(200)$ features on average.}
    \label{fig:fiducial}
\end{figure}
\paragraph{Choice of binning.} In the interest of an invertible covariance matrix and a reliable Fisher matrix, given that we have $120$ realizations in total per cosmology, we need a summary statistic with fewer than $120$ bins.  Choosing the optimal number of bins necessitates a compromise between two competing effects: with too few bins, we loose constraining power; with too many bins, the (inverse) covariance matrix is not reliable. (Additionally, for too many bins fluctuations in some bins may not be Gaussian.) As a trade-off between these effects, we use $30$ bins, corresponding to $5$ bins per distribution. We show an example of our data vector for the fiducial cosmology in Fig.\ \ref{fig:fiducial}.

\subsection{Fisher Matrix}
\begin{figure}
    \centering
    \includegraphics[width=1.0\textwidth]{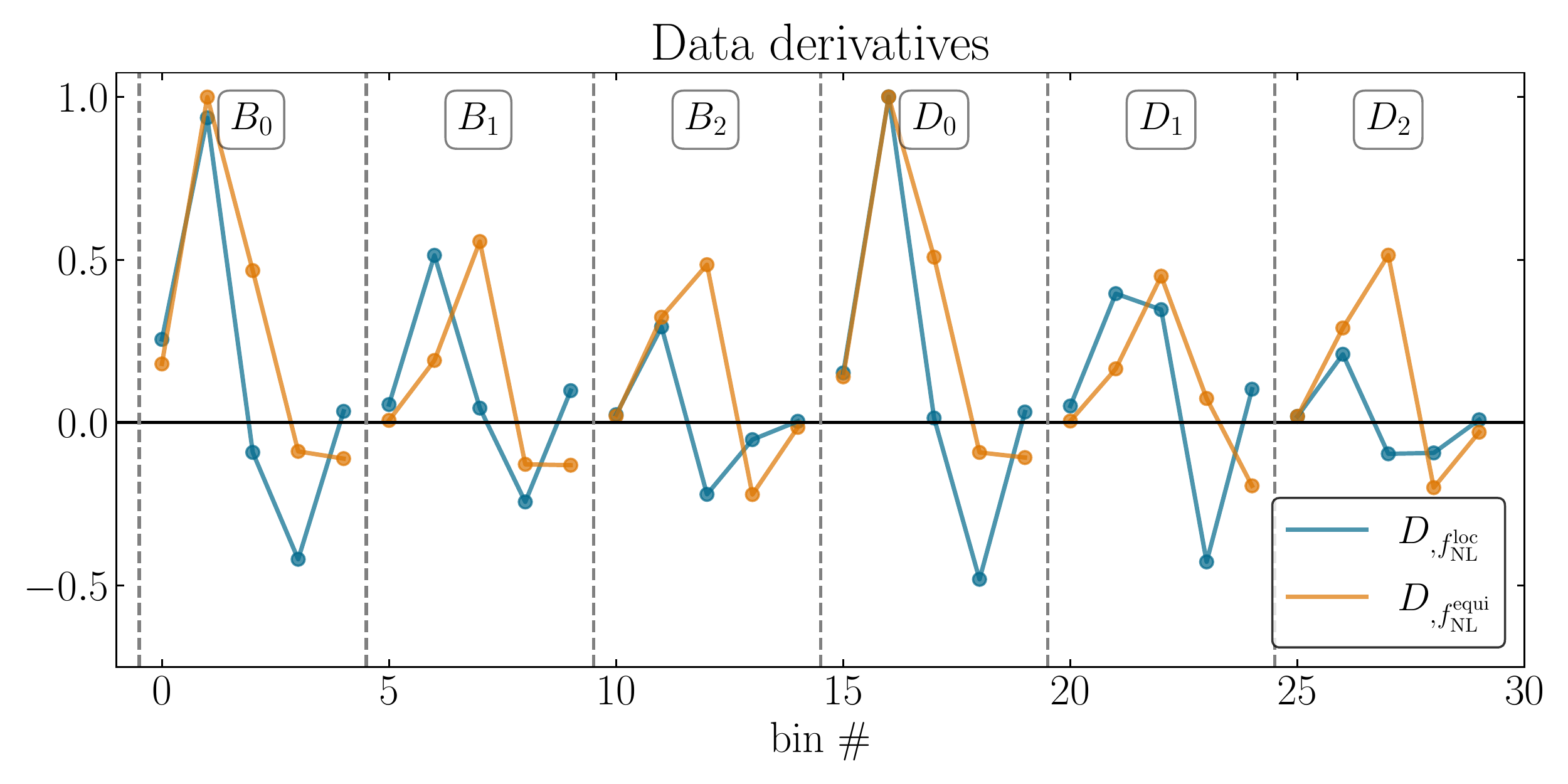}
    \caption{Numerical data derivatives, computed using $\fnlloc=10$ and $\fnleq=-30$. For visualization's sake, these are scaled to lie in the interval $[-1,1]$.}
    \label{fig:ders}
\end{figure}
In order to estimate the constraining power of our summary statistics, we compute the Fisher matrix. The Fisher matrix encodes the optimal constraints that can be derived from a summary statistic \cite{fisher1935logic,Tegmark:1998wm}. 
For Gaussian-distributed data, and neglecting the dependence of the covariance on the parameters,\footnote{In our setup, the covariance matrix actually depends on parameters significantly. One would then be tempted to add the term $\frac{1}{2}{\rm Tr}\left[\left(C^{-1}C_{,i}\right)\left(C^{-1}C_{,j}\right)\right]$ to $F_{ij}$, i.e. the contribution from the derivative of the covariance. However, if the mean and variance of our statistics are not independent, including this term can overcount some the information content of a statistic \cite{Carron:2012pw}.  Since our summary statistics are derived from discrete counts of features in bins, their fluctuations are closely related to those of Poisson statistics. For a Poisson distribution, the variance equals the mean, and the Fisher matrix is only given by the term in \eqref{eq:Fisher}.} the Fisher matrix is given by
\begin{equation}\label{eq:Fisher}
    F_{ij}=D_{,i}^T C^{-1}D_{,j}
\end{equation}
where $D$ is the data vector of size $N_b = 30$ bins, averaged over $15$ realizations each with $8$ sub-boxes for each cosmology, $X_{,i} = (X(\theta_i) - X(\theta_i=0))/\theta_i$ denotes the numerical derivative of $X$ with respect to model parameter $\theta_i=\fnlloc,\fnleq$. The covariance matrix $C$ is built directly from the data vectors,
\begin{equation}
    C_{ij} = \langle(D_i - D_{\rm mean})(D_j - D_{\rm mean})\rangle,
\end{equation}
where $i,j=1,...,60$ run over the elements of the data vector of the Gaussian initial conditions simulations, \textsf{G85L}, $D_{\rm mean}$ is the mean data vector over $15$ realizations and $8$ sub-boxes for a total of $N=120$, and $\langle \cdot \rangle = \frac{1}{N-1} \sum_{i=1}^N \cdot $ is the average over $N$. Since we use the data covariance matrix, when inverting we include a correction factor  $C^{-1} \longrightarrow \frac{N - N_b -2}{N-1}\, C^{-1}$ \cite{Hartlap:2006kj}. Given the Fisher matrix, the marginalized information on model parameter $\theta_i$ corresponds to a 1-$\sigma$ constraint of $\sigma_i = \sqrt{\left(F^{-1}\right)_{ii}}$.

\paragraph{Numerical derivatives.} For a reliable estimate of the numerical derivative, it is generally preferred to use small deviations of the reference parameter, assuming that for these the response of the quantity to the parameter is linear. We use $\fnlloc=10$ and $\fnleq=-30$ to compute numerical derivatives. For these amplitudes, the numerical derivatives of our summary statistics can be estimated reliably, i.e.\ they are not noise-dominated. We show convergence of our results with respect to the simulation volume used to estimate the derivatives and covariance in App.\ \ref{app:converge}. As observed in \cite{Biagetti:2020skr}, the change in our summary statistics scales nonlinearly with $\fnleq,\fnlloc$. In particular, the response scales as $D(\fnleq)-D(\fnleq=0)\sim (\fnleq)^n$ for $0 < n < 1$. See App.\ \ref{app:largefnl} for details. We show the numerical derivatives in Fig.\ \ref{fig:ders}.

\paragraph{Gaussianity tests.} True constraints will be most closely related to the Fisher estimate if fluctuations of our summary statistics are Gaussian.
We visually inspect our bins, drawing a histogram of counts for all the realizations given a bin and verifying that all bins are Gaussian distributed. 
Since each bin by definition includes a non-negative number of counts,
crucial to this approximation is that each bin of our distribution is sufficiently populated, namely that the mean is several times larger than the square root of the variance. Imagining Poisson noise, we would have that the variance scales with the mean. With 60 bins, we find that each bin has a mean of at least $4.5~\sigma$, while we find $10~\sigma$ for 30 bins. We decide to use the latter configuration.  As a further test of the Gaussianity, we perform a Kolmogorov-Smirnov test\cite{kstest} for this choice of bins. We find that our summary statistics passes the test well within the $95\%$ confidence limit.\footnote{We thank an anonymous referee for suggesting the test.}

\section{Results}\label{sec:results}

\begin{figure}
    \centering
         \includegraphics[width=0.49\textwidth]{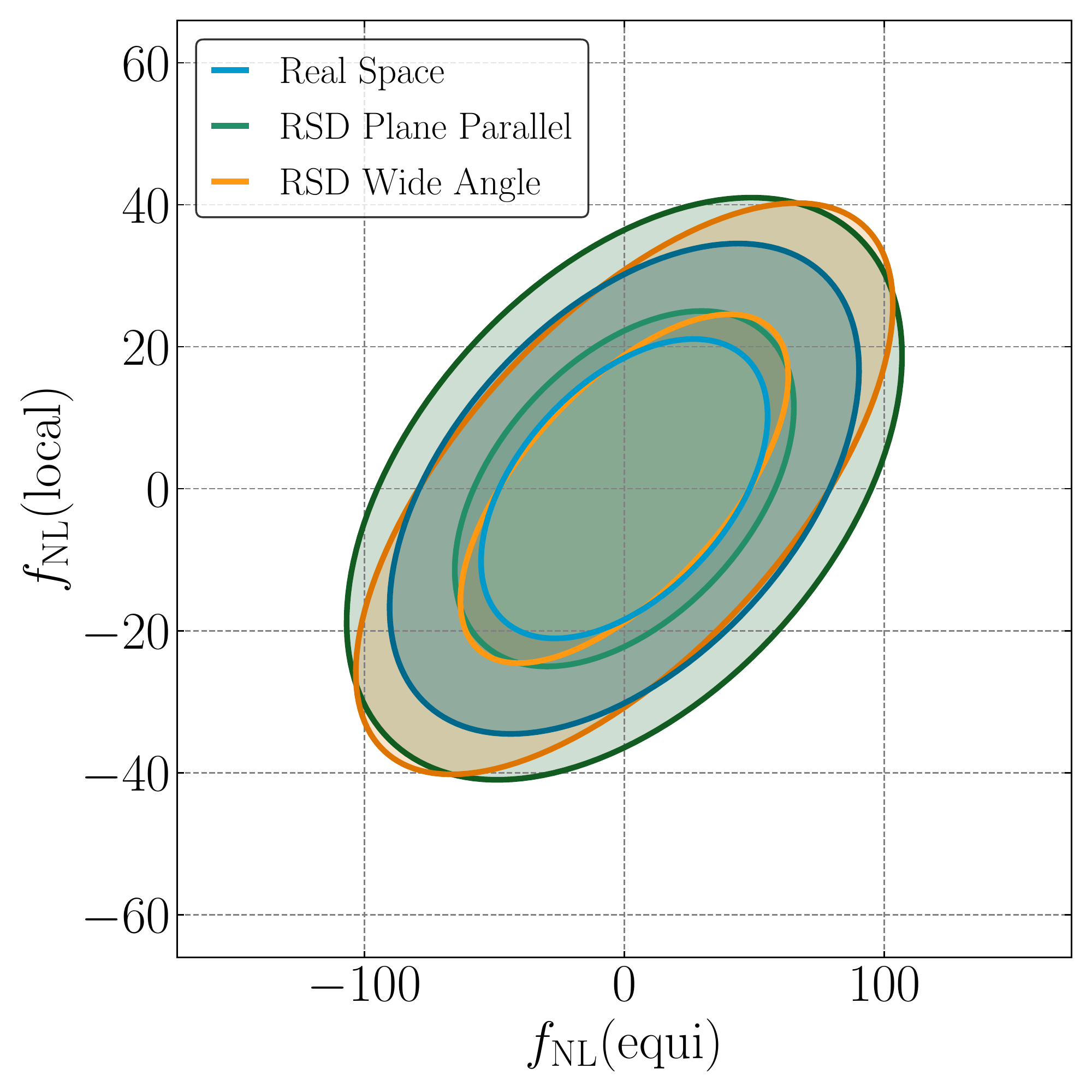}
         \includegraphics[width=0.49\textwidth]{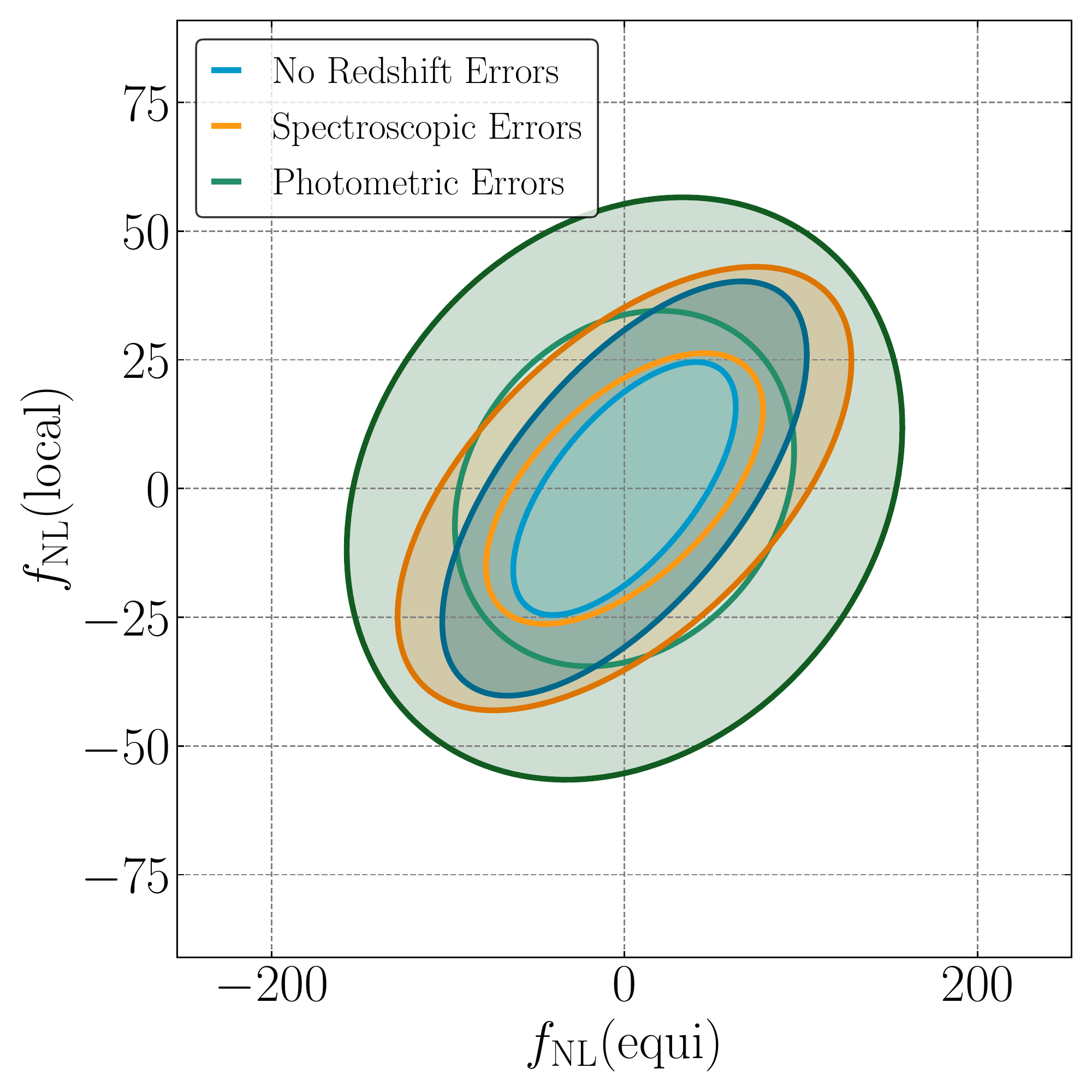}
        \caption{ \textbf{Fisher forecasts.} \emph{Left Panel:} Constraints in the $\fnleq-\fnlloc$ plane for real space (blue contours), redshift space in the plane-parallel approximation (green contours) and redshift space wide angle observer (orange contours). \emph{Right Panel:} Constraints in the $\fnleq-\fnlloc$ plane for the case of the wide angle observer in redshift space without redshift errors (blue) and with spectroscopic (orange) and photometric (green) errors.} 
        \label{fig:main}
\end{figure}

We present our main results in Figure~\ref{fig:main}. First, we show on the left panel the uncertainty on $\fnlloc$ and $\fnleq$ obtained from the Fisher matrix as described in the previous section, for a fiducial cosmology with Gaussian initial conditions. We present the results for the analysis in real space (blue contours), in redshift space in the plane-parallel approximation (green contours), and in redshift space without the plane-parallel approximation where the observer is in a corner of the simulation box (orange contours). 

The 1-$\sigma$ marginalized constraints are of the order of $\Delta \fnlloc \sim 15$, $\Delta \fnleq \sim 40$, with small differences among the three cases. The degeneracy between $\fnlloc$ and $\fnleq$ is seen to be small. As a consequence of this, the constraints on one of the two parameters do not degrade very much when marginalizing over the other. The redshift space analysis gives slightly weaker constraints, with the plane-parallel case being more pessimistic. 
The analysis in redshift space is straightforward to perform since we need to simply displace the halos' apparent position and run the pipeline on the displaced halos. In particular, the analysis in the wide-angle case is as easy as the plane-parallel case. This is an advantage over analyses based on the comparison of the measured power spectrum to the theoretical prediction based on perturbation theory, where computing wide-angle effects is not trivial \cite{Castorina:2019hyr}. 

\paragraph{Impact of redshift errors} We estimate the effect of photometric and spectroscopic redshift errors on this analysis, which are modeled as described in Sec.~\ref{sec:implementation}. We present these results in the right panel of figure~\ref{fig:main}. We present error contours for the case with no redshift errors, photometric errors, and spectroscopic errors. In all these cases the observer is in a corner of the simulation box. The results shown were obtained after averaging over six realizations of the errors.

Overall, the inclusion of redshift errors leads to a mild degradation of the constraints. As expected, photometric errors have the largest impact, degrading constraints by $\sim 50\%$ for $\fnleq$ and $\sim 25\%$ for $\fnlloc$. Spectroscopic errors only change the contours slightly, by $\sim 20\%$ for $\fnleq$ and $\lesssim 10\%$ for $\fnlloc$. All these results are summarized in Table \ref{tab:Real-space}. The protection against random perturbations of the points' location in the volume is a typical property of persistent homology, so we might say that these statistics are protected from redshift errors by construction. See Sec.\ \ref{sec:small} for more discussion on this point. It is also worth pointing out that redshift errors are instead a strong source of degradation for a joint power spectrum-bispectrum analysis \cite{Karagiannis:2018jdt}, especially for equilateral non-Gaussianity. 

\begin{table}[h]
\begin{centering}
\begin{tabular}{|c|c|c|c|c|c|}
\cline{2-5} \cline{3-5} \cline{4-5} \cline{5-6} 
\multicolumn{1}{c|}{} & Real Space & RSD Plane Parallel & RSD Wide Angle & $+ v_{\rm spec}$-error &$+ v_{\rm photo}$-error \tabularnewline
 \hline
$\Delta \fnlloc$ & 13.9 & 16.5  & 16.2 & 17.3 & 22.8\tabularnewline
\hline 
$\Delta \fnleq$ & 36.3 & 43.0 & 41.6 & 51.8 & 63.4\tabularnewline
\hline 
\end{tabular}
\par\end{centering}
\caption{{\bf Fisher forecasts.} Comparison of Real-Space results with Redshift space in the plane-parallel approximation without redshift errors and Redshift space wide angle observer results with and without redshift errors.  }
\label{tab:Real-space}
\end{table}

\begin{figure}[t]
    \centering
         \includegraphics[width=0.55\textwidth]{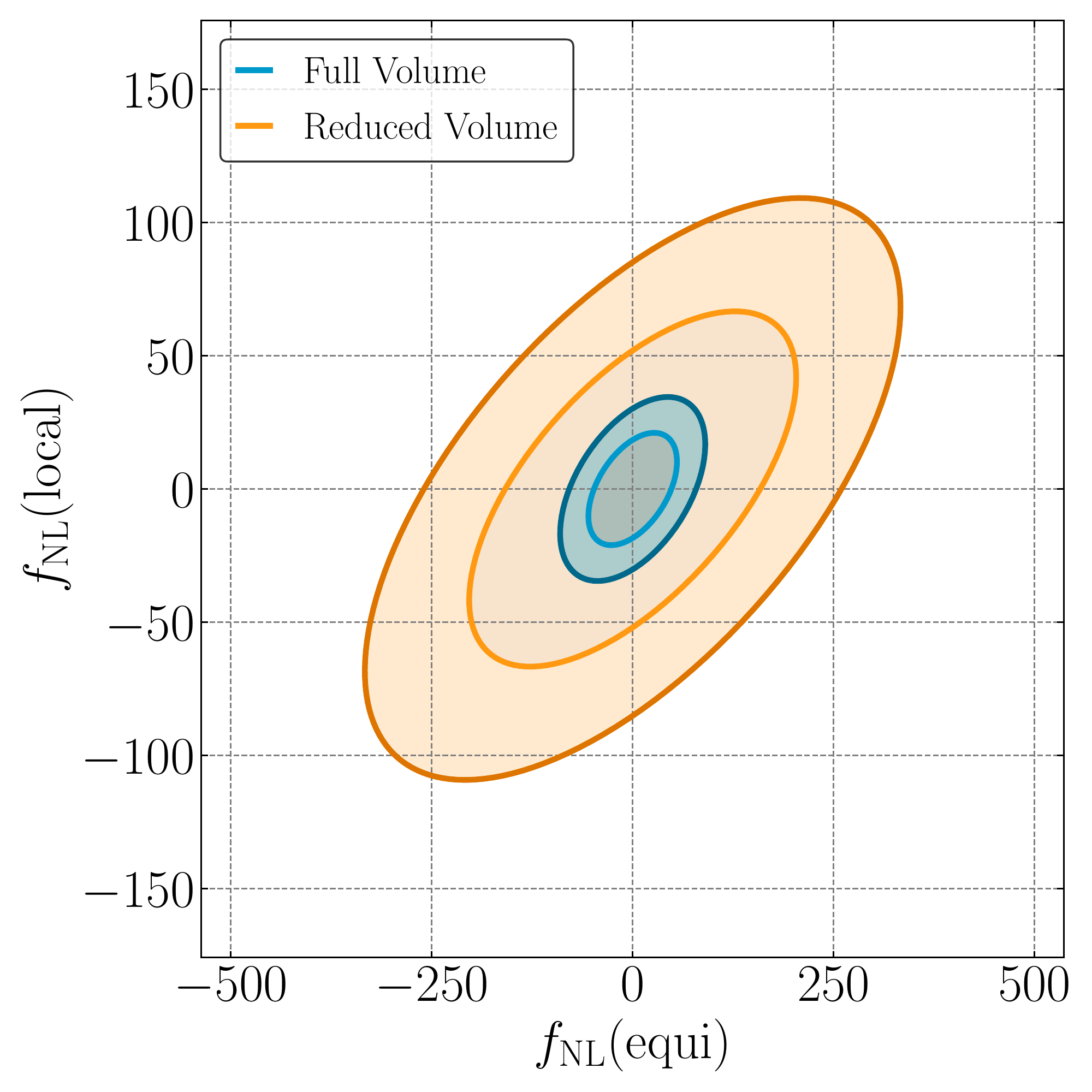}
        \caption{ \textbf{Fisher forecasts with reduced volume.} We compare the constraints obtained from a reduced volume (orange) to the ones already shown for the full one (blue). The reduced volume is $0.125$ $($Gpc$/h)^3$, which is $8$ times smaller than the volume considered in the previous forecasts.  }
        \label{fig:results_volume}
\end{figure}

\paragraph{Dependence on volume} We also wish to estimate how these constraints scale with the volume of the survey. For this, we repeat the real space analysis reducing the volume of the sub-boxes by a factor of $8$. This is done by dividing each full simulation box of $2$ Gpc$/h$ aside into $64$ sub-boxes of side $500$ Mpc$/h$. The results are presented in figure~\ref{fig:results_volume}. We see that this reduction of volume leads to a degradation of the contours by a factor of roughly $\sim 3$. This factor of $\sim 3$ is what one would expect from a Poisson-like distribution. In this case, we have that $D_{,i}\sim {\rm vol}$ and $C^{-1}\sim \frac{1}{\rm vol}$, so that $F_{ij}\sim {\rm vol}$. Then marginalized constraints scale as $\sigma \sim \sqrt{\left(F^{-1}\right)_{ii}}\sim \left({\rm vol}\right)^{-1/2}$. We have confirmed numerically that $D_{,i}\sim {\rm vol}$ and $C^{-1}\sim \frac{1}{\rm vol}$ for our simulations, so that  $\frac{\sigma_{500}}{\sigma_{1000}}\sim \sqrt{8}\sim 3$, in line with the scaling in Fig.\ \ref{fig:results_volume}. More precisely, we find $\Delta \fnlloc = 44.0$, $\Delta \fnleq = 134.1$. The fact that the scaling is slightly worse than expectation could be due to the fact that a smaller box has larger surface-area-to-volume ratio, so that some features are cut off by the box's boundaries and $D_{,i}$ scales slightly sublinearly with volume.
\begin{figure}[t]
    \centering
    \includegraphics[width=0.33\textwidth]{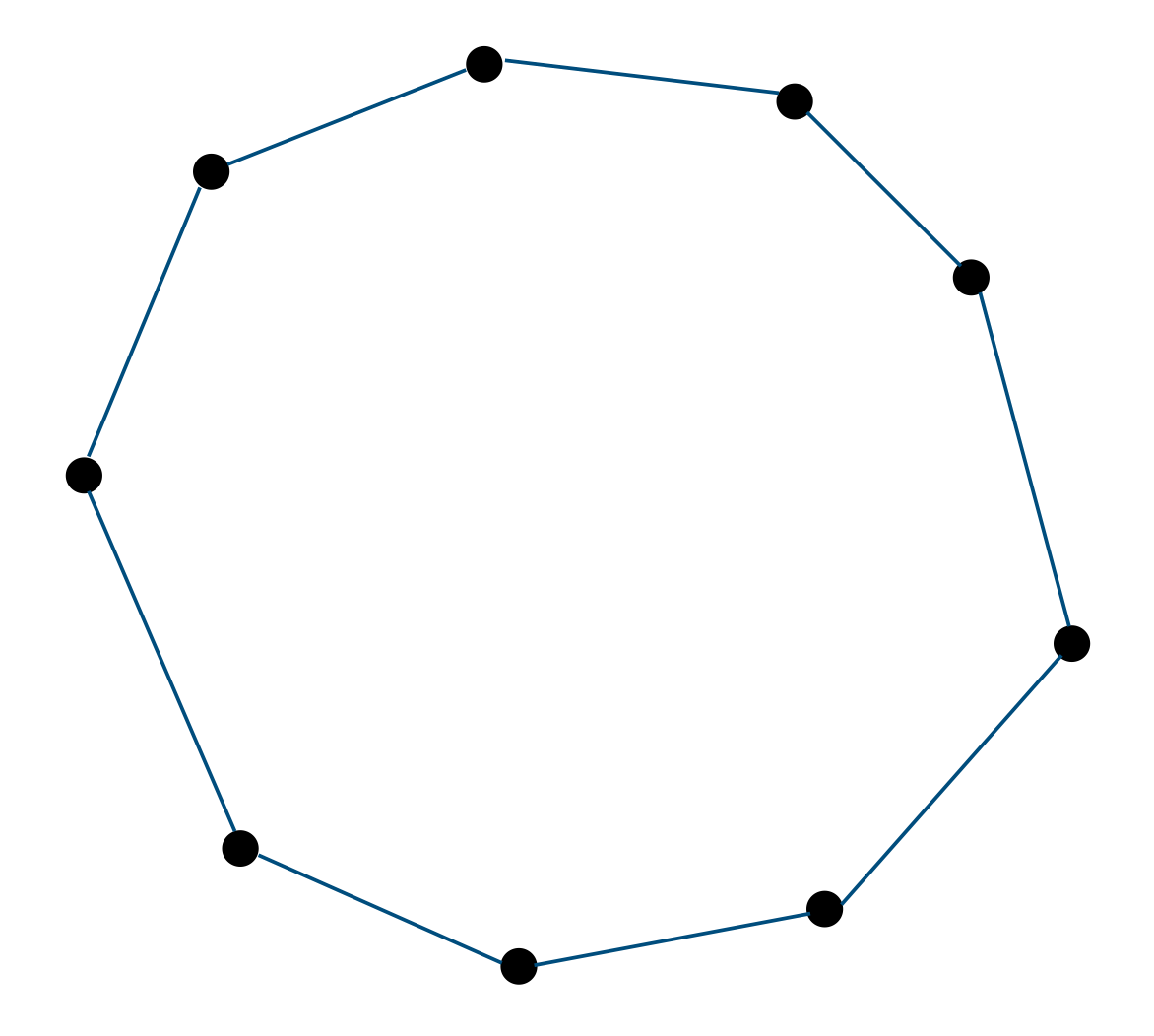}\includegraphics[width=0.33\textwidth]{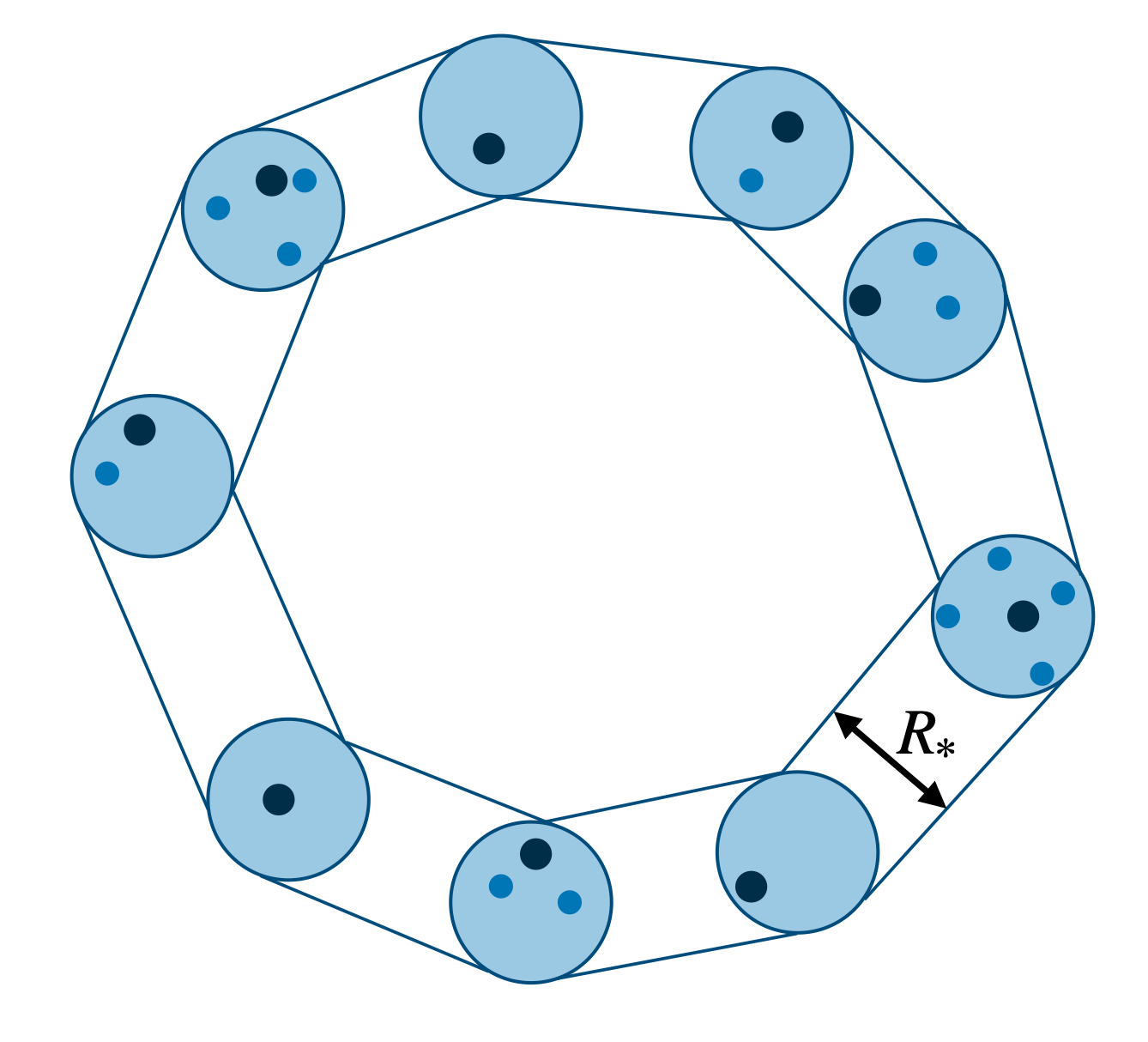}\includegraphics[width=0.33\textwidth]{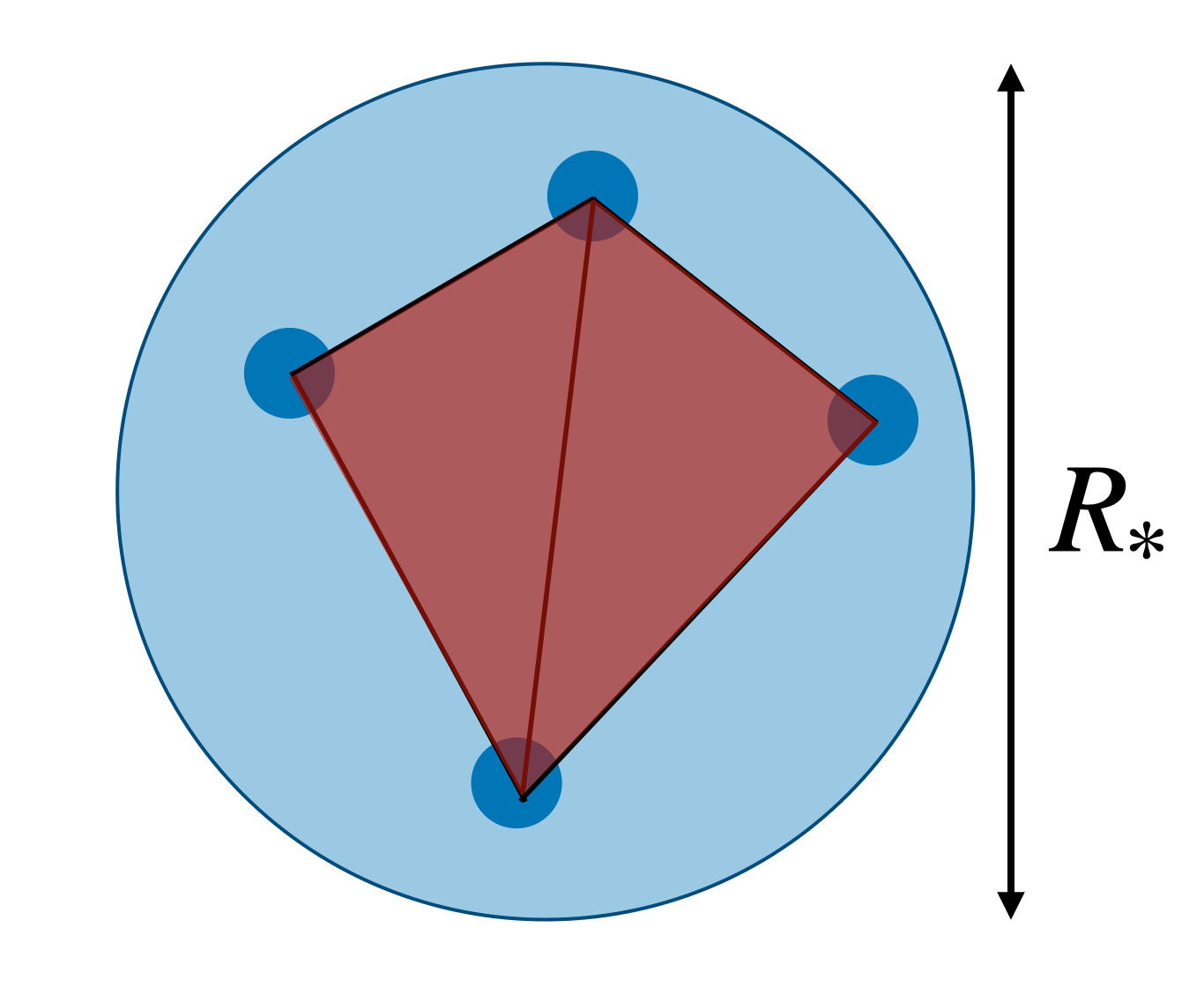}
    \caption{ \emph{Left Panel:} a filament loop before perturbations. \emph{Middle Panel:} A filament loop after perturbations. Initial halos are displaced within balls of radius $R_*$, and subhalos are also included. The birth and death scale of the filament loop change at most by $\mathcal{O}(R_*)$. \emph{Right Panel:} uncertainties may create topological features, but these are erased at a scale $\leq R_*$.}
    \label{fig:perturbations}
\end{figure}
\section{Comments on robustness against small-scale uncertainties}\label{sec:small}

We have seen that redshift errors do not significantly degrade our forecasted constraints on primordial non-Gaussianity.
In this section, we discuss more broadly the robustness of our topological statistics against general small-scale uncertainties. 
First of all, we should note that while a filtration is defined semi-locally, topological features are by definition non-local. Therefore, they are suited to tracing non-local statistics of the survey map. Consider local small-scale uncertainties parameterized by the scale $R_*$. For example, $R_*$ can be a scale at which gravitational nonlinearties (or galaxy formation) cannot be modeled by our simulation, as in \cite{Baumann:2021ykm}. Alternatively, $R_*$ could be the typical scale of redshift errors, as in the previous section. 

What impact can perturbations at the scale $R_*$ have on our topological statistics? At their most dramatic, these perturbations can create or destroy topological features. However, these features are necessarily parameterized by a scale $R<R_*$. For example, a loop parameterized by scale $R\gg R_*$ cannot be destroyed by a perturbation of scale $R_*$. On the other hand, we expect that the birth and death scale of this feature can change, but only to $\mathcal{O}(R_*)$. This can be regarded as a form of topological protection. This process is depicted in Fig.\ \ref{fig:perturbations}. 

Therefore the coarse-graining scale implemented by a filtration allows us to identify and cut topological features that may depend on small-scale uncertainties, for example ignoring feature that die below the scale $R_*$. Additionally, significant topological features may only be affected to order $R_*$, so that
under suitable binning the distributions of these features are robust.

Note that this intuitive argument is essentially a statement about the continuity of persistence diagrams under perturbations to the data. For simpler filtrations than ours, this statement can be formalized and proven as a ``stability theorem'' \cite{cohen2007stability}. From previous analyses of \eos \cite{MoradinezhadDizgah:2020whw}, we have that a conservative estimate for $R_*$ in our present context is the size of the largest halo, $R_*\sim 2-5$ Mpc/h. From Fig.\ \ref{fig:PD}, we see that all topological features are born and die beyond this scale. Therefore at the level of dark-matter-only simulations, small-scale dynamics is not expected to change our constraints. Note that a similar scale of perturbations is provided by redshift errors in eqns. (\ref{eqn:zerrors1},\ref{eqn:zerrors2}).
\begin{figure}[t]
    \centering
         \includegraphics[width=0.49\textwidth]{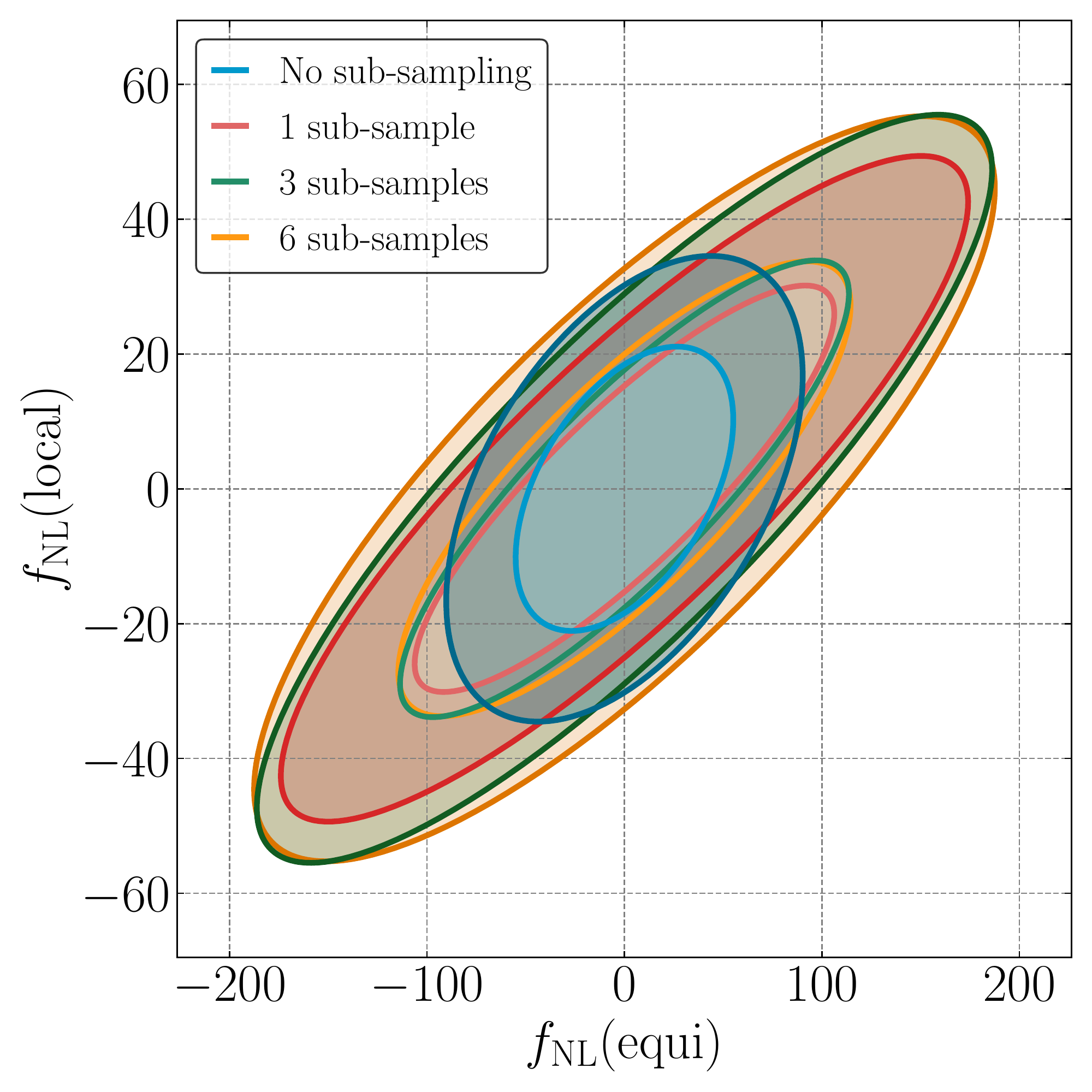}
        \caption{ \textbf{Fisher forecasts for sub-sampled data.} Constraints in real space for the case of not sub-sampled data (blue) and for sub-sampled data from $1$ realization (red), $3$ realizations and $6$ realizations.}
        \label{fig:subsample}
\end{figure}

The more pressing question comes when comparing N-body simulations to galaxy survey data, for which the dynamics by which galaxies form and populate halos is important.
As an example for small-scale uncertainties in our present context, we can think of a typical Halo Occupation Distribution (HOD) algorithm, which describes how halos are occupied by galaxies. In HOD models, low-mass halos may or may not be populated by any galaxies, while for high-mass halos one may have not only central galaxies but also ``satellite'' galaxies, depending on a set of parameters. The situation for high-mass halos is not worrisome due to the previous argument. The deletion of low-mass halos is slightly different in nature. Since most halos are in fact low-mass, this can have the effect of significantly changing the number density of tracers, which can affect the scales at which topological features form and disappear. Additionally, we might expect that deleting a halo can have significant effect on nearby topological features.

To estimate the effect of such deletions, we perform a procedure similar to \cite{Biagetti:2020skr}, where each halo catalog was randomly sub-sampled so that data derivatives were computed at fixed number density. Considering that most halos are at low-mass, this procedure should, at least qualitatively, mimic the HOD effect on central galaxies we just discussed. As a test on the current setup, we therefore sub-sample the halo field by imposing that all sub-boxes across all cosmologies have the same number density. This amounts to randomly removing halos up to $\mathcal{O}(5)\%$ over the total number. Since this procedure is random, we perform it several times with a different random seed. We show this effect in figure \ref{fig:subsample}. Sub-sampling degrades the forecasted constraint on $\fnleq$ by a factor of $\sim 2$. On the other hand, constraints on $\fnlloc$ are more robust, degrading by about $\sim 25\%$. The degeneracy is also increased. We run several realizations of the random sub-sampling. We observe that the constraints converge for more than $3$ realizations. 
These results call for a more careful quantitative analysis of these uncertainties using a calibrated HOD implementation. Additionally, we might expect that a smart choice of filtration would mitigate the effect of deleting some points. We leave this to future work.

\section{Conclusions}\label{sec:conclusions}

We have presented a class of summary statistics that trace information regarding primordial non-Gaussianity at the map level. Our Fisher estimates demonstrate that significant information can be extracted by considering higher-order structures in position space. The particular structures we exploited are topological in nature, describing clustering, filament loops, and voids across cosmological scales. We forecast $\Delta \fnlloc \sim  16$ and $\Delta \fnleq \sim 41$ using a volume of $\sim 1$ $($Gpc$/h)^3$ in redshift space without the distant observer approximation. Our results are quite robust to uncertainties in the determination of the redshift of each halo. Contours degrade by $\lesssim 20\%$ when considering spectroscopic errors, and by $\lesssim 50\%$ when considering photometric errors.

These results show a strong potential for primordial non-Gaussianity as compared to recent analyses. For instance, a Fisher forecast using the halo power spectrum and bispectrum in real space on the \eos simulations considered in this work gave $\Delta \fnlloc\sim20$ using full boxes of $8\,($Gpc$/h)^3$ volume with a realistic theoretical covariance \cite{Biagetti:2021tua}. As for real data constraints, the power spectrum and bispectrum as measured from the BOSS survey was recently analyzed in the context of primordial non-Gaussianity and gave constraints of the order of $\Delta\fnlloc\sim 50$ and $\Delta\fnleq\sim 200$ at $68\%$ confidence on a volume of $2.4$~$($Gpc$/h)^3$~\cite{DAmico:2022gki}.

There are several avenues forward. For one, we have presented in this work expected constraints based on a Fisher matrix formalism. These implicitly Gaussianize hand-crafted summaries of our topological statistics. While our hand-crafted summaries are well-approximated by Gaussian distributions, they do not necessarily convey the entire information content of a persistence diagram. The tradeoff is that the likelihood for our persistence diagrams is only implicitly defined. Parameter estimation in the context of implicit likelihoods is precisely within the purview of the rapidly advancing field of simulation-based inference (\cite{greenberg2019automatic,papamakarios2019sequential,hermans2020likelihood,Miller:2021hys}, see \cite{cranmer2020frontier} for a recent review and \cite{Alsing:2019xrx,Villaescusa-Navarro:2021cni,Villaescusa-Navarro:2021pkb,Makinen:2021nly,Cole:2021gwr} for cosmological applications). Within simulation-based inference it has been advocated (see e.g.\ \cite{Alsing:2019xrx}) to perform data compression in two steps: first by constructing by hand an interpretable summary statistic and then by allowing a neural network to extract the maximum information content from that statistic. In our context, we would eventually like to tune our topological summaries for manifest robustness against small-scale effects. In some sense, this is a form of nuisance-hardening \cite{Alsing:2019dvb}.
 
Ultimately, precise robustness against small-scale physics such as HOD must be verified numerically. This will require the development of a new simulation suite that includes both primordial non-Gaussianity and HOD. We may also speculate as to whether the physical principles elucidated in \cite{Baumann:2021ykm} would allow for fast and cheap N-body simulation strategies that directly target accuracy at the length scales where primordial non-Gaussianity can be directly distinguished from late-time effects. At any rate, the summary statistics introduced in this work demonstrate that significant information regarding primordial non-Gaussianity can be extracted in a manifestly position-space approach. Although we cannot yet demonstrate robustness against every conceivable small-scale effect, we anticipate that when considering more realistic models, the position-space definition and topological nature of our statistics will be very helpful.

Additionally, as we ultimately aim to make contact with observation, there are several complications that must be incorporated, including window functions, survey geometry, and more. We look forward to reporting on progress in this direction in future work.

\section*{Acknowledgments}
We thank Daniel Baumann, Dan Green and Pierluigi Monaco for fruitful discussion on small-scale non-linearities. We thank Cora Uhlemann for useful comments and suggestions on a draft.
M.B.\ acknowledges support from the Netherlands Organization for Scientific Research (NWO), which is funded by the Dutch Ministry of Education, Culture and Science (OCW) under VENI grant 016.Veni.192.210. M.B.\ also acknowledges support from the NWO project ``Cosmic origins from simulated universes'' for the computing time allocated to run a subset of the \eos simulations on \textsc{Cartesius}, a supercomputer which is part of the Dutch National Computing Facilities.  Additionally, a subset of the \eos simulations on \textsc{Cartesius} and its successor \textsc{Snellius} were run under the project ``Topological echoes of primordial physics in cosmological observables.'' A.C.\ was partially funded by the Netherlands eScience Center, grant number ETEC.2019.018.  J.C. is supported by ANID scholarship No. 21210008. L.C. acknoledges support from  ANID scholarship No.21190484 and ``Beca posgrado PUCV, t\'ermino de tesis, 2021''. J.N. is supported by FONDECYT grant 1211545, “Measuring the Field Spectrum of the Early Universe”.

\appendix
\section{Consistency checks}
\subsection{Convergence tests}\label{app:converge}
\begin{figure}
    \centering
         \includegraphics[width=0.49\textwidth]{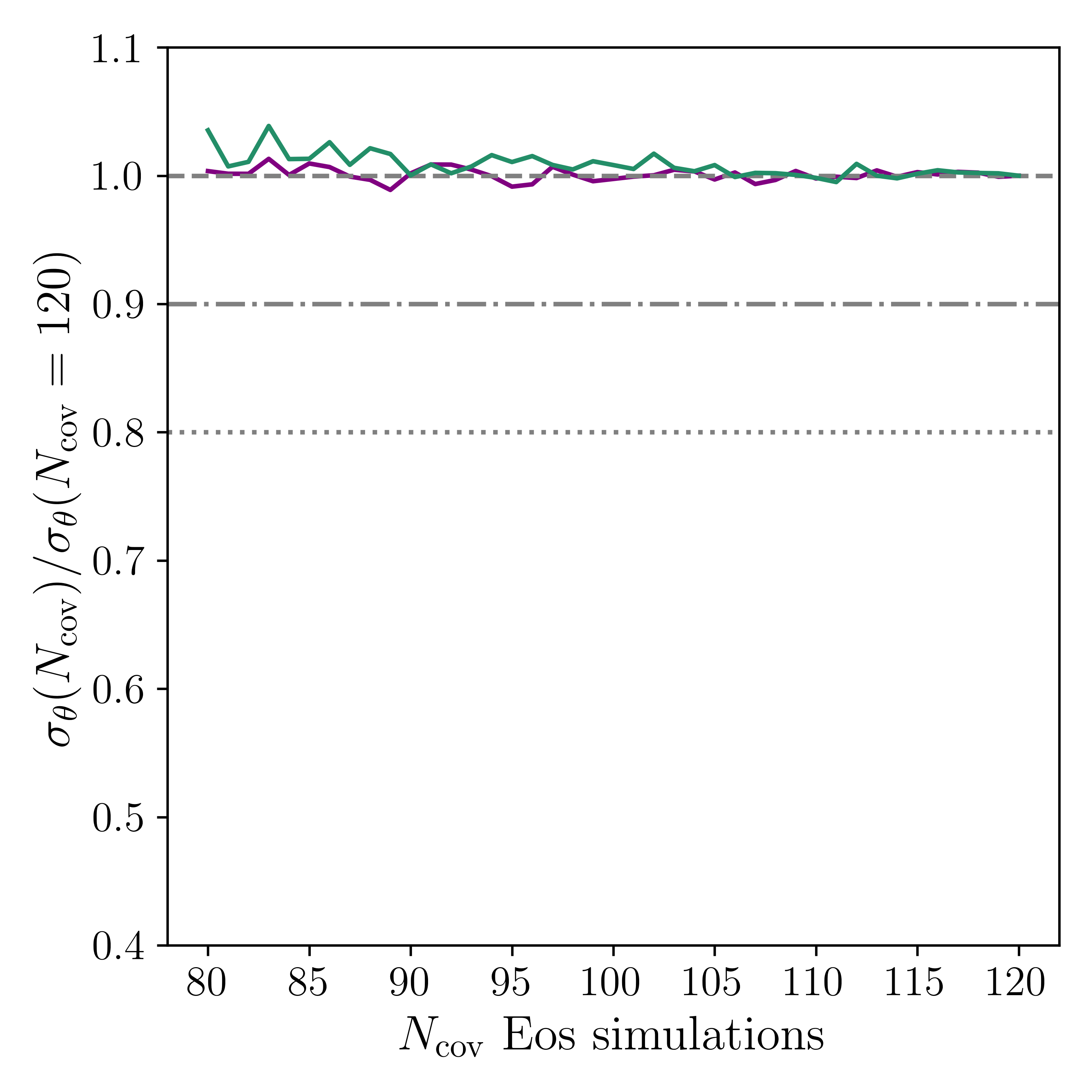}
         \includegraphics[width=0.49\textwidth]{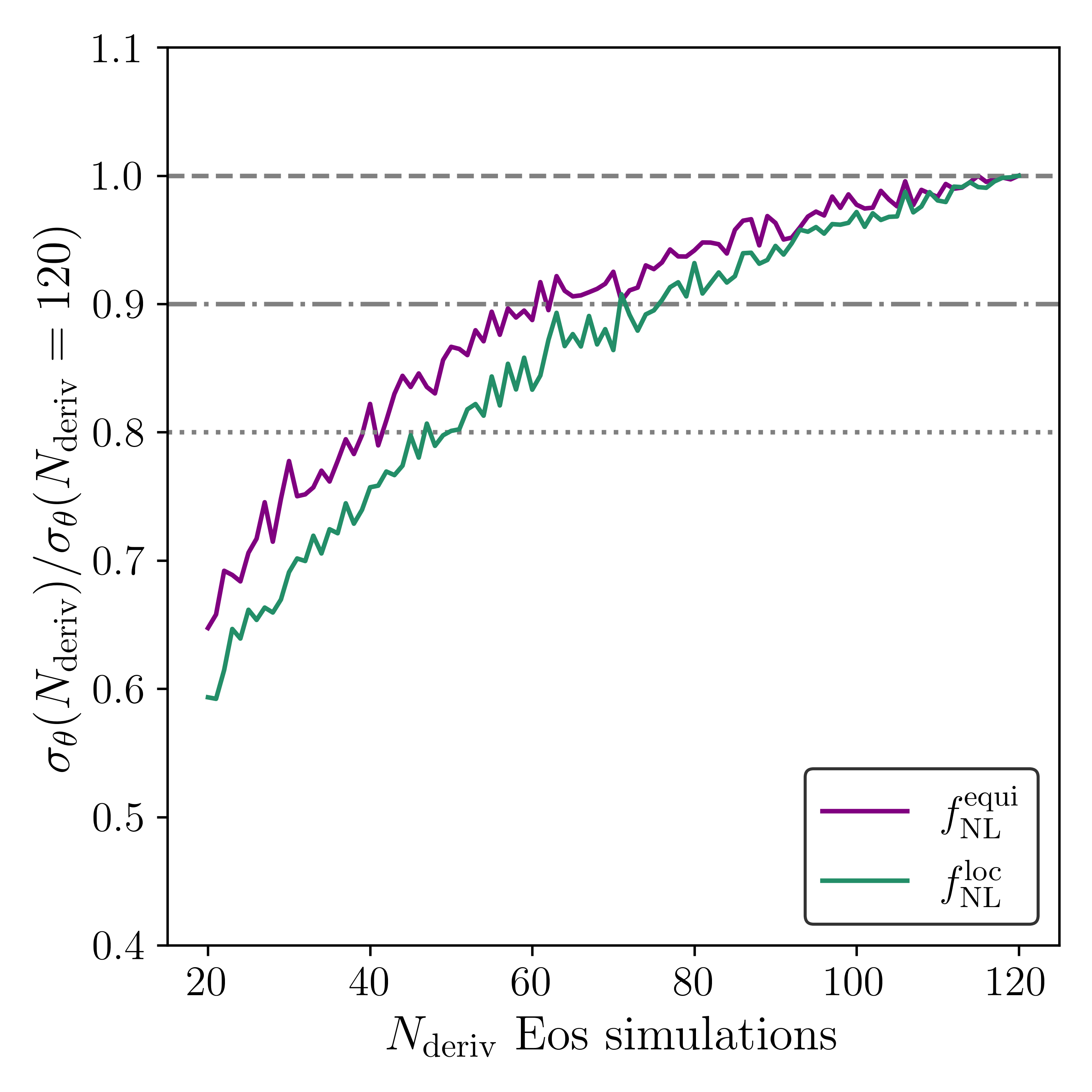}
        \caption{ {\bf Convergence tests for halo catalogs in real space}. \emph{Left Panel}: convergence of  constraints with respect to the number of simulations used to compute the covariance matrix. \emph{Right Panel}: convergence with respect to number of simulations used to compute numerical derivatives. Shown are averages over 50 random orderings of the simulations.} 
        \label{fig:totconv}
\end{figure}

Since we estimate the components of eq.\ \eqref{eq:Fisher} numerically, we must check that our results are converged with respect to simulation volume. 
As a heuristic for convergence, as in \cite{Hahn:2019zob}, we may examine the ratio of the uncertainty $\sigma_{\theta}(N)$ on each parameter $\theta$ for varying number $N$ of realizations over the uncertainty $\sigma_{\theta}(N=120)$ for the maximum number $N=120$ of realizations. We consider convergence with respect to the number of simulations used for estimation of the covariance matrix $N_{\rm cov}$ and data derivative $N_{\rm der}$ separately. 
In Figure \ref{fig:totconv} we show the convergence of marginalized constraints. We see that fluctuations with respect to $N_{\rm cov}$ affect the forecasted constraints within a few percent. On the other hand, and as in \cite{Hahn:2019zob}, $\sigma$ grows monotonically with $N_{\rm der}$. However, this growth flattens, and at $60\%$ of the available simulation volume the results remain within $10\%$ of their final values. This signals that if we had access to more simulation volume, the final results would not change by very much.
We can heuristically extract extrapolated constraints via the ansatz $\sigma(N) = \sigma_\infty - 2^{- (N-N_0)/\tau_n}$. Performing this check, we find that our forecasts are stable within $20\%$.

\subsection{Large Non-Gaussianity}\label{app:largefnl}
\begin{figure}
    \centering
         \includegraphics[width=0.49\textwidth]{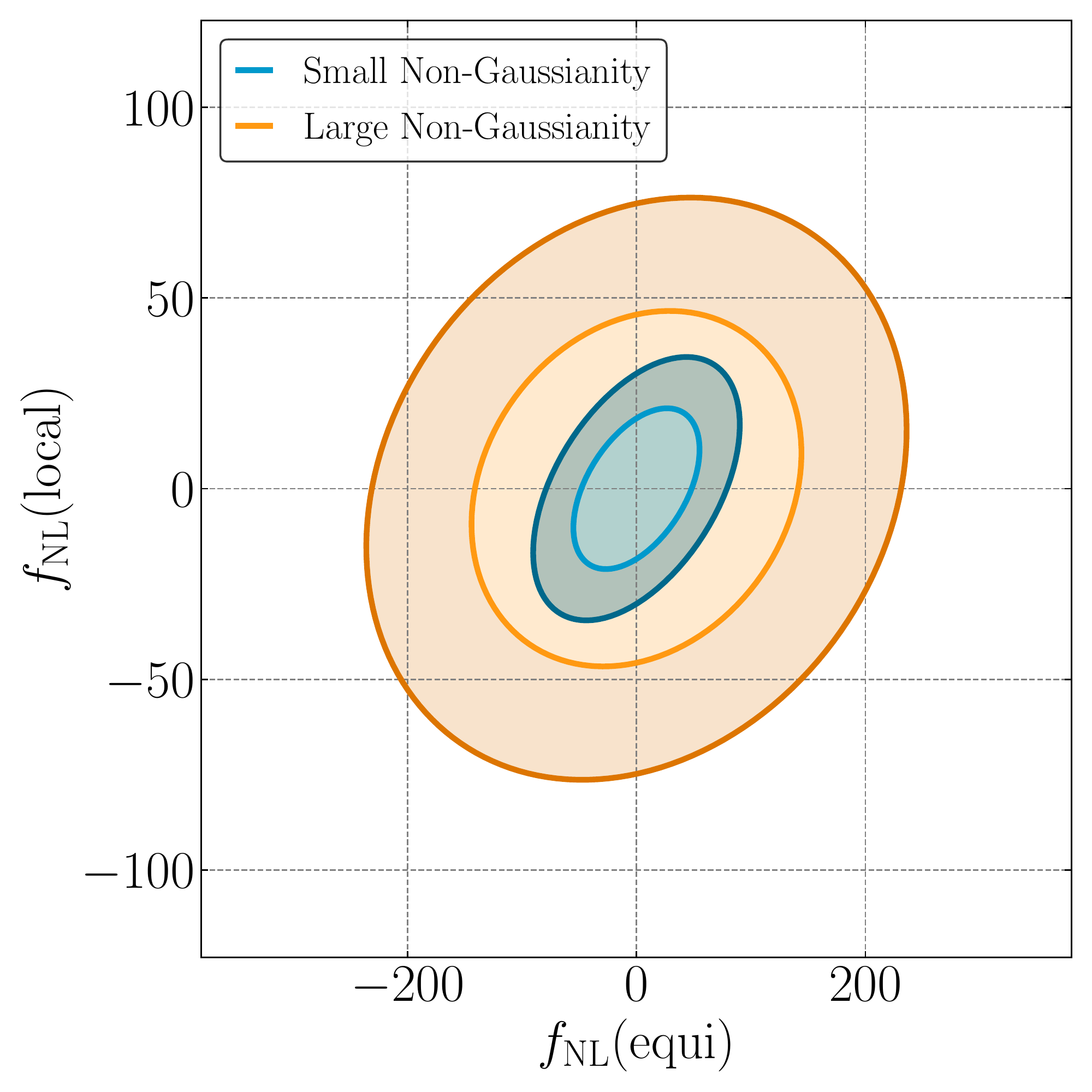}
        \caption{ \textbf{Fisher forecasts with large $f_{\rm NL}$.} Constraints in real space for small (blue) and large (orange) non-Gaussianity. The sizes of $f_{\rm NL}$ used to estimate derivatives are $\fnlloc=10$ and $\fnleq=-30$ for the small non-Gaussianity case and $\fnlloc=250$ and $\fnleq=-1000$ for the large non-Gaussianity case.}
        \label{fig:largefNL}
\end{figure}
As observed in \cite{Biagetti:2020skr} in the case of local non-Gaussianity, the non-Gaussian contribution to topological statistics scales nonlinearly with $f_{\rm NL}$. In particular, the scaling appears to be sublinear, so that using large values of $f_{\rm NL}$ to estimate the data derivative underestimates the constraining power of our topological summaries. On the other hand, since the signal is quite large, these results converge more quickly with respect to simulation volume. We show the Fisher ellipse when data derivatives are estimated using $\fnleq=-1000$ and $\fnlloc=250$ in Fig.\ \ref{fig:largefNL}. We observe that marginalized constraints degrade by a factor of $\sim 2.7$. We note the consistency of this factor via the following investigation. First suppose that the scaling of the difference in data vector goes as $(d_f-d_0)\sim f^{x}$. In this case, $x$ can be computed by performing the minimization $\textrm{argmin}_y ||(d_F-d_0) - y(d_f-d_0) ||$. Rearranging, one has that $x=\frac{-\ln(y)}{\ln(f/F)}$. Here the final result depends on the convention for the norm $||.||$. When using $C^{-1}$ to compute the norm, more attention is given to bins with smaller variances. Using a Euclidean norm, more attention is given to bins with larger means. We observe that the lower variance bins are more dramatically sublinear. For definiteness, consider local non-Gaussianity. Using $C^{-1}$ for the norm, we find $y\sim 6$, while for a Euclidean norm, we find $y\sim 18$. There is a similar behavior when comparing $\fnleq=-30,-1000$. These lead us to expect that $\sigma_{250}/\sigma_{10} \in [1.4, 4.2]$ -- consistent with the results in Fig.\ \ref{app:largefnl}. We anticipate that more detailed modeling of the behavior of our topological statistics will enable greater simulation-efficiency in future studies.

\bibliographystyle{utphys}
\bibliography{Halos-TDA}

\end{document}